\documentclass{jpp}
\usepackage{url}
\usepackage{hyperref}
\usepackage{graphicx}
 \usepackage{epstopdf, epsfig}
\usepackage[caption = false]{subfig}
\usepackage[utf8]{inputenc}
\usepackage[T1]{fontenc}
\usepackage{amsmath}

\author{Mangilal Choudhary, Vanshika, Surya}
\affiliation{Department of Physics and Astrophysics, University of Delhi, Delhi-110007, India}
\title{Transient spark dielectric barrier post-discharge plasma reactor with a liquid electrode for dye degradation: A primary study}
\begin{document}
\maketitle
\begin{abstract}
The potential application of non-thermal plasma in treating textile industrial wastewater motivates researchers to develop innovative techniques at a laboratory scale to achieve the goal of wastewater mineralization. In line with this objective, a dielectric barrier post-discharge plasma reactor with a liquid electrode has been built for the study of synthetic dye degradation. The plasma reactor was optimized by altering various operating conditions to achieve a higher degradation efficiency at given discharge conditions. The reaction kinetics of crystal violet degradation were studied, and the same plasma reactor was tested for other synthetic dyes (wastewater model samples). The results suggest that the proposed dielectric barrier post-discharge plasma reactor may offer a promising solution for treating dye effluents from the textile industry.
\vskip 2mm
\end{abstract}
\maketitle
\maketitle

\section{Introduction}
Mitigating and controlling environmental pollution, which includes air pollution, soil pollution, and water pollution, has become a significant challenge for the future. The presence of pollutants in the air, water, and soil has a negative impact on the health of living organisms and contributes to climate change. Numerous reports indicate that industrial effluents and untreated sewage are significant contributors to water pollution. The textile and dyeing industry effluent contains various types of toxic organic dyes that are assumed to be hazardous to the environment if untreated effluents are discharged into water bodies \cite{toxicitydyebook2,effectdyeolivingorganism_19}. Therefore, the removal of organic synthetic dyes from industrial wastewater is an important issue in environmental research. However, a wide range of work around the globe has been ongoing to develop advanced techniques for wastewater that contains dyes and organic compounds \cite{physicalchemicalmethode2, physicalandchemicalmethod1, wastewaterallmethodes,  chemicalmethod2, biologicalmethod1,biologicalmethod2,biologicalmethod3}.\\
There are some challenges in removing highly stable organic dyes from wastewater using physical, biological and chemical processes. The physical techniques (e.g., filtration, adsorption, flotation) are generally efficient, but they only transfer dissolved pollutants (organic compounds or dyes) from the liquid phase to the solid phase. Therefore, post-treatment of the solid wastes and regeneration of the adsorbent material are required.   
\cite{physical_methods_review, adsoptionphysical_6,nanofiltrephysical_4,nanofilterphysical_5}.\\ 
The advanced oxidation and photocatalytic techniques, which are chemical processes, have been investigated for the degradation of synthetic organic dyes in aqueous solutions. The strong oxidizers (mainly ozone - $O_{3}$, hydrogen peroxide - $H_2 O_2$, hydroxyl radicals ($\dot{OH}$, atomic oxygen -$O$) interact with organic pollutants (e.g. complex dye molecules, organic molecules) and decompose them into less toxic molecules \cite{advancedoxidation1,advancedoxidation2,advancedoxidation3,advancedoxidation4,advancedoxidation5}. Sometimes these strong oxidizers are coupled with catalysts \cite{ozonationwithcatalyst, fentonmethodsdyedegrade_2012} and catalysts are exposed by UV radiations \cite{photodegradtaionreview2024} to make advanced oxidation techniques more reliable for organic pollutants degradation. It is noted that the use of chemicals, the synthesis of catalysts, and high energy consumption make chemical advanced oxidation techniques more costly and have an adverse environmental impact.   \\ 
The experimental reports suggest that biological treatment can be effective in place of chemical methods for the removal of organic dyes or molecules from wastewater \cite{microbialdegradation_2017, microbialdyedegradtaion_2013, microorganisndegrdation2021,akansha2019decolorization,biologicaldye_muthu2022_book, biologicalmethod3}. But due to the high stability of organic molecules, especially dyes, biological processes are sometimes ineffective in degrading all kinds of complex organic dyes or molecules from water. It refers that some cost-effective and eco-friendly advanced oxidation processes, along with biological and chemical treatment techniques, can be promising solutions for industrial wastewater mineralization (i.e. conversion of organic compounds to carbon dioxide, inorganic ions, water, etc.). \\\\  
In the last few years, electric discharges (non-thermal plasma) have also been investigated as an advanced oxidation technique for the degradation of various organic compounds, including synthetic dyes \cite{paracetamoledegradeplasma-water_2024, wasterwaterplasma1, grabowski_2007_mb_pulsedcorona, wastewaterplasma2, organicdye_diaphramdischarge_2009}. The Non-thermal plasma-based advanced oxidation techniques can generate various strong oxidizing elements (e.g. UV photons, $O_3$, $\dot{OH}$, $H_2O_2$ and other oxygen reactive species) without using any external chemicals, which helps in degrading the organic compounds in water \cite{plasmawaterinteractionreview1,rosspecieswater1,wastewaterplasma4}. The various electric discharge (non-thermal plasma) configurations, such as pulsed corona discharges \cite{mbdegradtaionpulseddbd_2009, grabowski_2007_mb_pulsedcorona, ns-pulsed-dyedegradation_ao_2013, nspulsedcorona_dydegradation_2019}, direct current (DC) diaphragm discharge \cite{organicdye_diaphramdischarge_2009}, transient spark discharges \cite{rons_formation_spark_air_discharge_2013, vyas2023degradationmethylenebluedye}, atmospheric plasma jet \cite{plasmaject_dyedegrade_2015, plasmajet_dyedegrade_satya_2022, plasmajet_rathore_dyedegrade_2024}, gliding arc discharges \cite{glidingarc_dyedegradation_2006, glidingarc_dyedegradation_2012}, surface dielectric barrier discharges \cite{surfacebarrier_dbd_dyedegrade_2022}, corona needle-plate configuration \cite{needle-plate_dyedegradation_2016, multiholes_dbd_dyedegradation_2020}, dielectric barrier discharges \cite{dbd_waterdroping_dyedegrade_2013,wire-cylinder_post_dbd_dyedegradation_2014}, and  post-discharge or gas phase dielectric barrier discharges \cite{gasphasedbd_submerged_dyedegrdation_2008, postdischarge_gasphasedbd_dyedegradtaion_2015, post-discharge_dbd_dyedegradation_2029} are investigated to decompose the organic compounds and dyes in wastewater. \\\\
It is a fact that ozone is a long-lived, strong oxidizing species in the gaseous phase of non-thermal (corona or dielectric) plasma if air or oxygen is used as the input gas. However, other reactive oxidizing species, such as hydrogen peroxide and hydroxyl radicals, are produced in the water phase during plasma-water interaction. The higher organic pollutants degradation rate can be achieved by maximum utilisation of reactive species, especially ozone. In the above-mentioned electric discharges (plasma jet, corona DBD, gliding arc, surface DBD, etc.), plasma in the gas phase interacts with a water solution (wastewater), and chemical reactions occur on the surface of the water solution. The limited area of the plasma-water interaction reduces the efficiency of such DBD plasma reactors in removing organic contaminants on a large scale \cite{vyas2023degradationmethylenebluedye}. It is possible to increase the interfacial or interaction area for the gaseous phase of plasma (post-discharge plasma) by dispersing oxidizing species in the wastewater in the form of bubbles. The dissolved ozone directly or indirectly reacts with organic compounds or dye molecules, decomposing them through an oxidation mechanism. For the practical application of DBD plasma reactors, further studies on post-dielectric barrier discharge plasma reactors are needed to achieve maximum organic compound degradation efficiency. By keeping this broad view in mind, a transient spark DBD plasma reactor is built and optimized to achieve the maximum dye degradation rate with suitable experimental conditions.            
\\\\
In this study, numerous experiments were conducted using a newly built transient spark DBD reactor to explore its efficiency in degrading organic pollutants (synthetic dyes) from different model wastewater samples (dye solutions). The crystal violet (Fisher make) dye was used to optimize the post-discharge DBD reactor to achieve maximum dye degradation rate. Later, the same DBD reactor was tested to degrade other organic dyes (Methyl Blue, Reactive Red 120, Methyl Orange). The degradation mechanism and reaction kinetics were investigated to gain a deeper understanding of the observed results.\\
The full experimental setup and instruments used in the present study are discussed in Sec.~\ref{sec:sec2}. Experimental methods and procedures are discussed in Sec.~\ref{sec:sec3}. The experimentally observed results with crystal violet (CV) dye are presented in Sec.~\ref{sec:sec4}. The kinetics of CV degradation is presented in Sec.~\ref{sec:sec5}. The experimentally observed results are discussed in Sec.~\ref {sec:sec6}. Sec.~\ref{sec:sec7} discusses the degradation of other dyes with this newly developed DBD reactor. A summary of the work and future directions is provided in Sec .~\ref {sec:sec8}.  
\section{Experimental setup and measuring devices} \label{sec:sec2}
\begin{figure}
    \centering
    \includegraphics[width=0.95\linewidth]{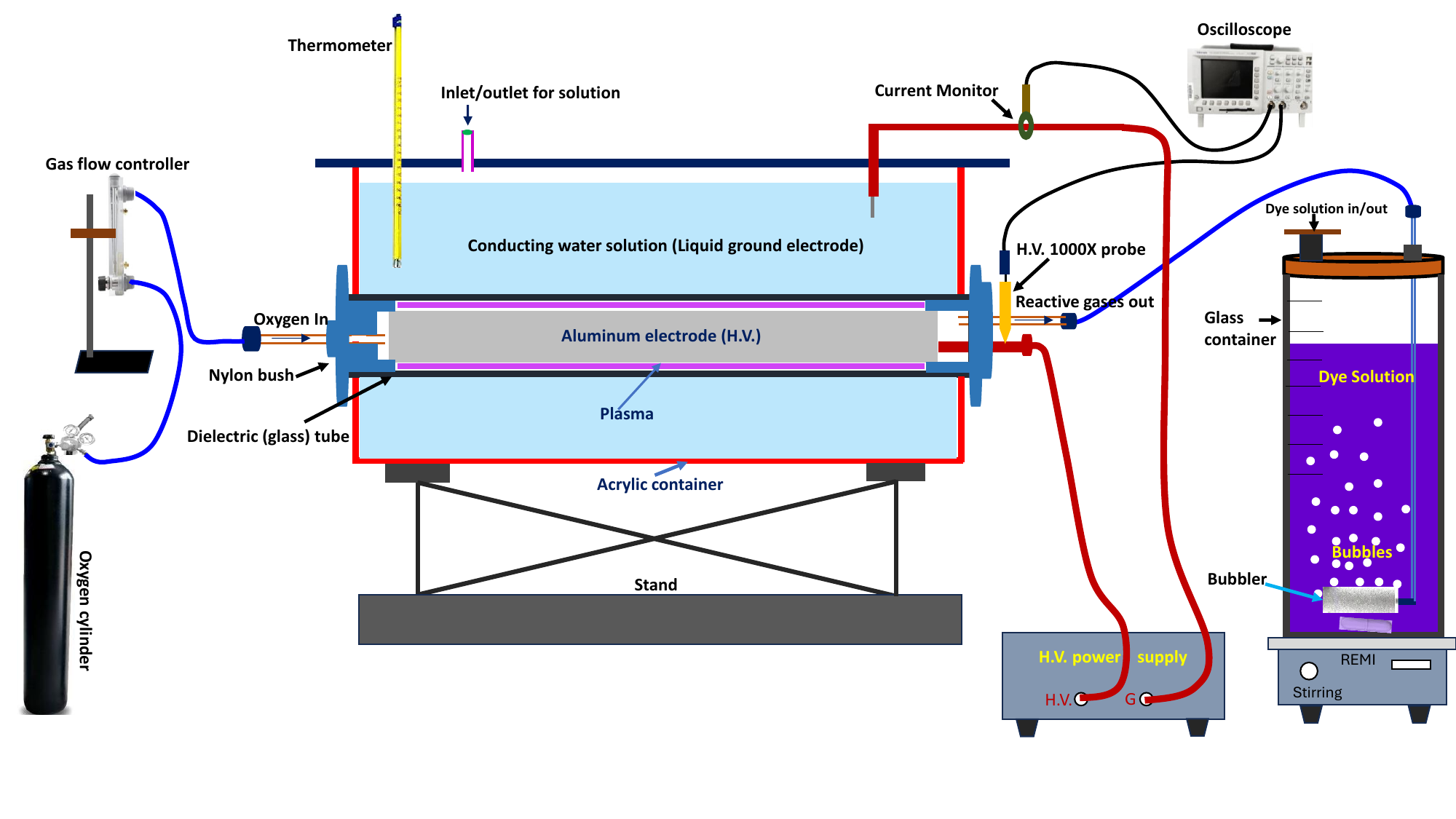},
    \caption{\textit{}{Schematic diagram of post-dielectric barrier discharge plasma reactor for degrading synthetic dyes (model wastewater samples).}}
    \label{fig:fig1}
\end{figure}

An in-house, low-cost dielectric barrier post-discharge plasma reactor was designed and built in the Plasma Physics Laboratory at the Department of Physics and Astrophysics, University of Delhi. A schematic diagram of the newly built plasma setup is shown in Fig.\ref{fig:fig1}. A borosilicate glass tube of outer diameter 22 mm, inner diameter 19 mm and length 200 mm was fixed in a rectangular acrylic container (as per Fig.\ref{fig:fig1}) of length 140 mm, width 90 mm and height 40 mm. An aluminium cylindrical rod of length 120 mm and diameter 16 mm was fixed at the centre of the glass tube through a nylon bush. Gas connectors were fitted to the nylon bushes for oxygen injection and the take-out of reactive gases from the glass tube. A gas flow rotameter (2 L/min) was used to control the input oxygen flow rate to the glass tube (discharge zone). An aluminium electrode is connected to the high-voltage (anode) terminal of the power supply, and the grounded terminal of the supply is inserted into the acrylic container filled with a conducting water solution. 5 g of sodium chloride was dissolved in 500 $ml$ of normal water to prepare a conducting water solution. Thus, the conducting water solution served as both a grounded electrode (cathode) and a coolant in this cylindrical dielectric barrier discharge (DBD) configuration. An in-house-built high-voltage (50 Hz, $V\sim$10 kV, max current 40 mA, 50 Hz) spark discharge power supply was used to breakdown the gases in the space between the aluminium electrode and the dielectric medium (glass tube wall). The discharge voltage and current were measured by using a high-voltage 1000X probe (Rigol RP108H) and a current transformer (Pearson 2877), respectively. The reactive gases were passed through a PU (polyurethane) pipe to the gas diffuser (gas bubbler). The gas bubbler was fixed into the prepared dye solution (model samples) close to the bottom of the glass container (see Fig.\ref{fig:fig1}). The glass container of 600 $ml$ (dye solution column) was placed on the magnetic stirrer (REMI 2L) to stir the dye solution during experiments. \\\\
The scientific-grade dyes, Crystal violet (Fisher make), Methyl blue (Otto make ), methyl orange (SRL make), and reactive red 120 (SRL make) are procured from authorized vendors. The temperature and pH of dye solutions were measured using a pH meter. Another meter provided the values of electrical conductivity (EC) and total dissolved solids (TDS) of the dye solution during experiments. A UV-Vis spectrometer (PerkinElmer Lambda 950) ) was used to monitor dye degradation percentage by measuring the height of the characteristic absorption peak at a given wavelength. The concentration of ozone ($O_3$) present in the reactive gas in the post-discharge configuration is measured using the Iodometric titration technique.
\section{Experimental methods and measurements} \label{sec:sec3}
\begin{figure*} 
 \centering
\subfloat{{\includegraphics[scale=0.230050]{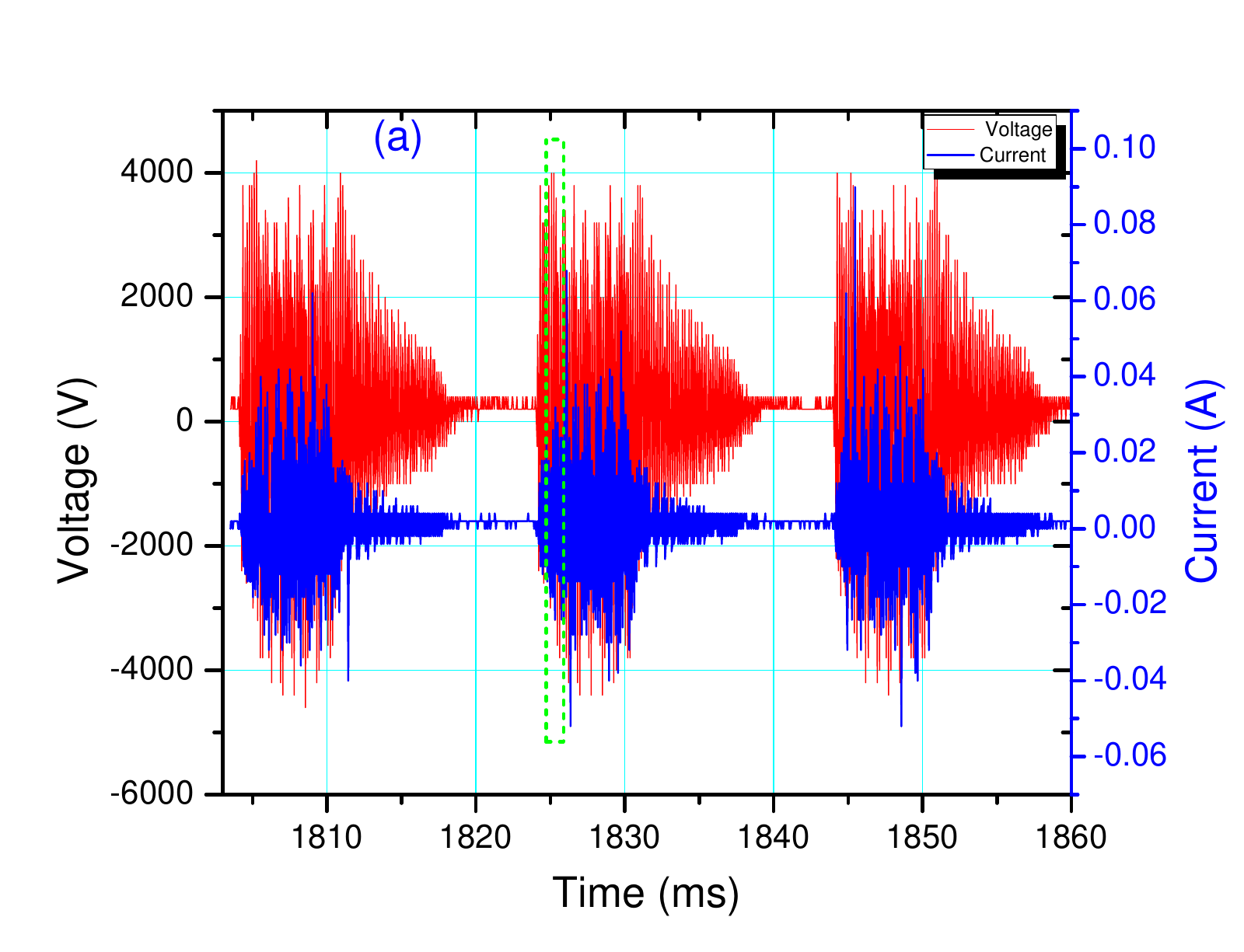}}}%
\subfloat{{\includegraphics[scale=0.23050]{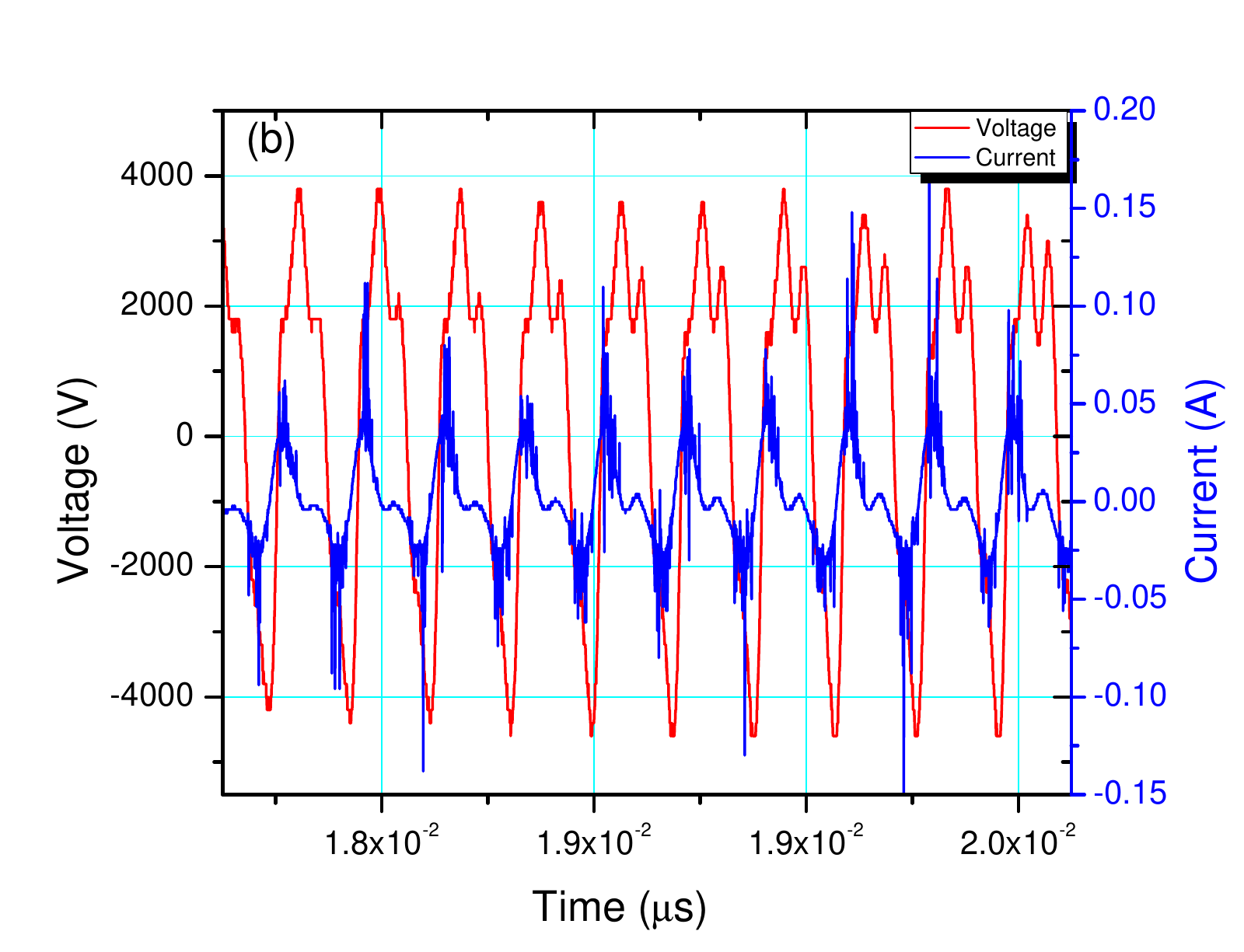}}}
\quad
\subfloat{{\includegraphics[scale=0.23050]{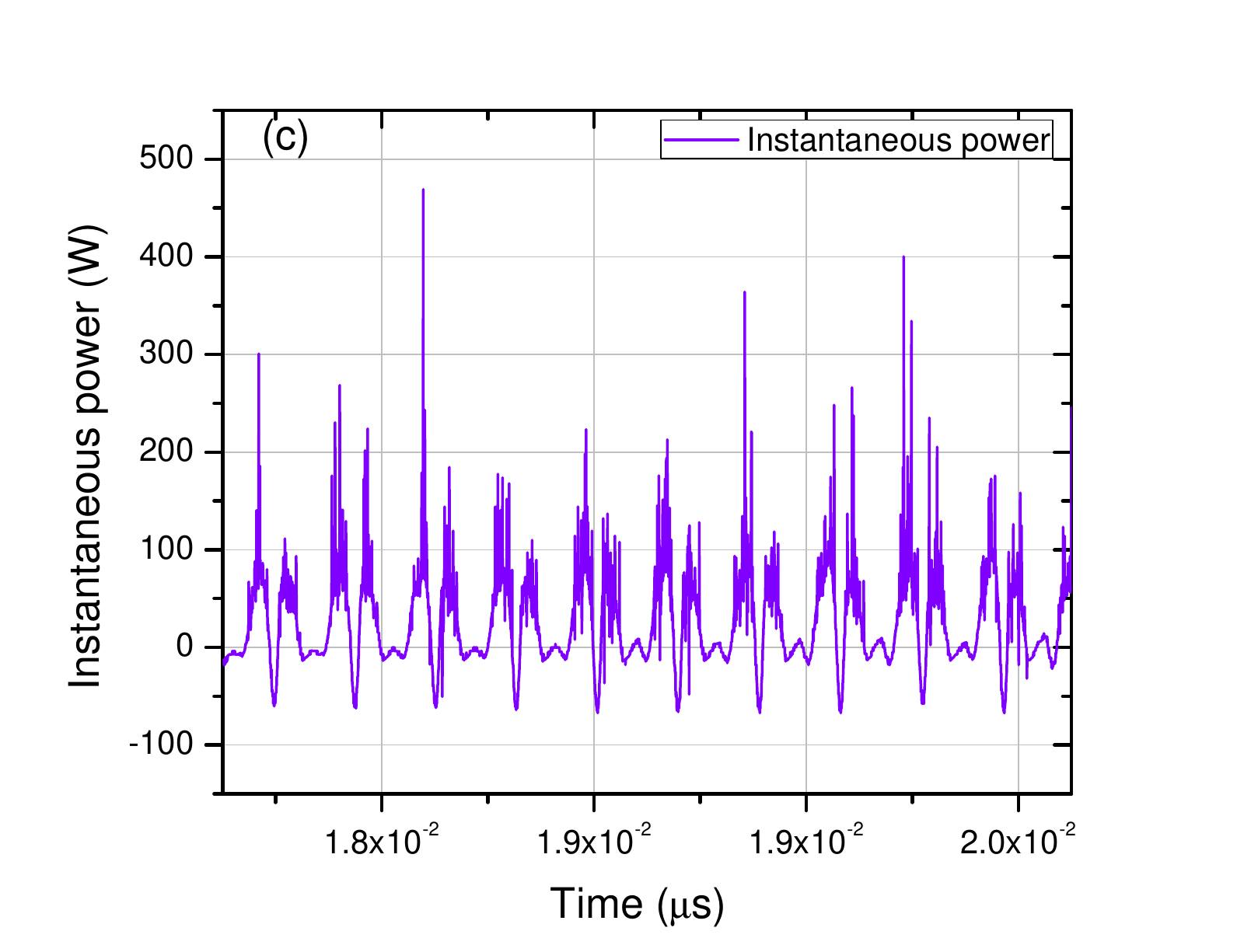}}}
\caption{\textit{} (a) Voltage and corresponding current waveforms at 50 Hz input signal. (b) Voltage and corresponding current waveforms at microsecond time scale (zoom view of the green dotted area in Fig.\ref{fig:fig2a}(a)). (c) Instantaneous power waveforms corresponding voltage and current signals in Fig.\ref{fig:fig2}(b).} 
\label{fig:fig2}
\end{figure*}

The DBD breakdown occurs in the gap between the dielectric and the anode (aluminium electrode) surface once AC high voltage is applied to the aluminium electrode with respect to the grounded liquid cathode. The duty cycle of the applied AC voltage signal is $\sim$ 50\%, and the frequency is 50 Hz. It means gas breakdown occurs only in the negative cycle of the input voltage signal. It is also possible to control the ON and OFF times of the applied voltage to the electrode by adjusting the timer settings. In the present work, we set 5 seconds ON time and 2 seconds OFF time to avoid damaging the electronics. It should be noted that only the on-time of the total interaction time of gas-liquid is considered as plasma treatment time in our calculations. Fig.\ref{fig:fig2}(a) shows the pattern of applied voltage at the aluminium electrode and corresponding discharge current during plasma formation. The maximum potential difference across the air gap is $>$ 8 kV, which is considered a high value for gas breakdown at atmospheric discharges. We observed sharp spikes in voltage and corresponding current due to the formation of micro-discharges between
electrodes. The presence of multiple current
Sharp spikes in microseconds (Fig.\ref{fig:fig2}(b)) confirm the repeated breakdown of oxygen in the discharge gap, which results in the formation of filamentary discharge channels. The zoomed view of the blue marked region in Fig.\ref{fig:fig2}(a) is presented in Fig.\ref{fig:fig2}(b). The instantaneous power, calculated as the product of $V(t)$ and $I(t)$, corresponding to Fig.\ref{fig:fig2}(b), is depicted in Fig.\ref{fig:fig2}(c). The estimated average power loss in one cycle (50 Hz) of output is estimated between 15 to 20 W. \\
The synthetic dye solutions were prepared in distilled water by weighing the dye powder on a weighing machine (Sartorius CPA64). Firstly, a high-concentration dye solution was prepared in distilled water, and then it was used to prepare dye solutions of different concentrations. During experiments, we always kept the lights dim to reduce the effects of photo-degradation. The prepared dye solution of the given concentration was  transferred to the experimental glass container (see Fig.\ref{fig:fig1}). As we apply a transient high voltage across the aluminium electrode and the grounded liquid electrode, gas breakdown occurs at a set oxygen flow rate, and a DBD plasma with short-lived filaments (micro-discharges) is formed between the metal electrode and the dielectric surface (tube wall). These filaments (micro-discharge) are essentially low-current plasma channels; therefore, various nitrogen and reactive oxygen species (both short-lived and long-lived) are formed in the discharge zone at atmospheric pressure. The long-lived reactive species, along with oxygen and other gases, are passed to dye solutions through the PU pipe and allow them to dissolve in the dye solution through the gas diffuser (bubbler). The degradation rate was monitored by analysing the UV-Vis spectrum of samples taken at different treatment times under various experimental conditions. Other parameters, such as temperature, pH, TDS, and EC, were also measured for both plasma-treated and untreated samples.   
\begin{table*} \label{table:table1}
\caption{Degradation of crystal violet dye in solution prepared with tap water and distilled water of different dye concentrations. Error over the estimated values (data) is within $<$ 5\%.}
\label{table:table1}
\centering 
\begin{tabular}{|l|c|c|c|c|c|c|c|c|c|c}
\hline
\multicolumn{5}{|c|}{Tap water dye solution} &
\multicolumn{5}{|c|}{Distilled water dye solution} \\
\hline
Con. & pH & EC & TDS & degradation time & Con.& pH & EC & TDS & degradation time  \\
 (mg/L)  &  & ($\mu$ S/cm) & (ppm) & (min) & (mg/L) & &($\mu$ S/cm) & (ppm) & (min)\\
\hline
0 & 7.83 & 325& 162.4 & -- & 0 & 7.03 & 0.86 & 0.43 \\
20& 7.85 & 328& 163.6 & 4.6 & 20& 6.88 & 3.26& 6.43 & 4\\   
 40 & 7.91 & 330 & 164 & 8.2 & 40 & 6.86 & 5.73 & 11.45 & 7.2\\
  60 & 7.93 & 333 & 154 & 11.5 & 40 & 6.89 & 8.35 & 15.75 & 10 \\   
\hline \hline
\end{tabular}
\end{table*}
 %
\section{Experimental observations}  \label{sec:sec4}
The major objective of this primary study was to optimize the DBD reactor in post-discharge configuration for degrading the synthetic dyes used in textile industries. The various sets of experiments were planned to investigate the impact of different experimental parameters on the decomposition of synthetic dyes. The crystal violet (CV) dye solution was taken as a model wastewater solution to optimize the DBD post-discharge configuration in this primary study. A detailed discussion on individual results is presented in the respective subsections.     
\subsection{Effect of the conductivity of the liquid cathode}
\begin{figure}
    \centering
    \includegraphics[width=0.65\linewidth]{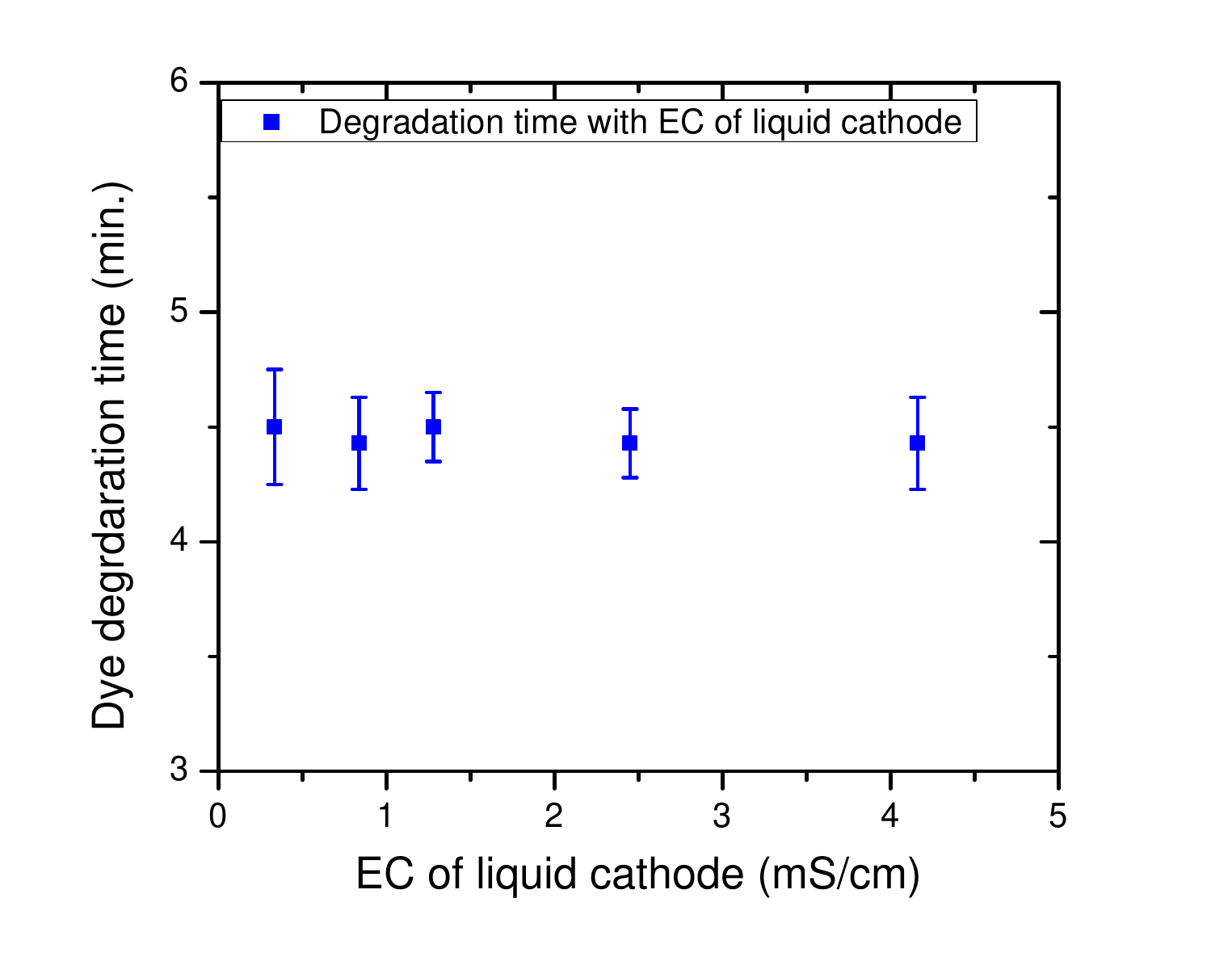},
    \caption{\textit{} Variation of degradation time for CV (at which dye degraded $>$ 98 \%) with changing conductivity of cathode liquid electrode. Volume of dye solution was 500 ml, dye concentration was 40 mg/L, gas flow rate was 400 cc/min, and dye solution temperature was 27\textdegree C.}
    \label{fig:fig3}
\end{figure}

Operating large-volume cylindrical DBD discharges with solid metal cathodes (mesh or metal sheets) is a challenging task due to the heating issue that can damage the reactor. Therefore, external cooling is necessary to operate the DBD reactor for an extended period to treat wastewater. To overcome this heating issue, we replaced the metal sheet (cathode) wrapped around the dielectric tube with a conducting liquid (cathode), as shown in Fig.\ref{fig:fig1}. The dielectric surface is in contact with the electrical conducting water solution, which provides a conducting path during discharge as well as cooling to the dielectric tube. In such a configuration, additional external cooling to the DBD reactor is not necessary to operate it for an extended period. The impact of the electrical conductivity (EC) of the liquid solution, which is working as a liquid cathode in the DBD configuration, on the CV degradation efficiency was investigated. Crystal violet solutions of similar concentrations were exposed to reactive gaseous species emanating from the DBD reactor by altering the EC of the cathode liquid. The time of dye degradation (solutions become transparent when more than 96\% degradation occurs) is plotted against the EC of cathode liquids in Fig.\ref{fig:fig3}. The rate of degradation is independent of the EC of the cathode liquid, as per experimental results (Fig.\ref{fig:fig3}). It clearly demonstrates that the characteristics of DBD plasma remain unchanged when the EC is varied from its reference value. Such information is essential to optimize the DBD reactor used in post-discharge configuration for achieving an efficient degradation rate.   

 \subsection{Role of the length of the exhaust pipe}
 In this DBD reactor, reactive gas species are transferred from the DBD reactor to the dye solution through a PU pipe. First, we tested the role of the length of the PU pipe. By reducing the length othe exhaust PU pipe by up to 30 cm, we did not observe a significant change in the degradation time of the f 40 mg/L CV solution. This knowledge enables the accurate determination of the exhaust pipe length without compromising the degradation efficiency in the post-discharge configuration.   
\subsection{Role of the DBD plasma volume}
The volume of DBD plasma depends on the contact area of the dielectric material (glass tube) surface with the liquid cathode. Experiments with a CV solution of 40 mg/L at a gas flow rate of 400 cc/min were conducted to investigate the effect of DBD plasma volume on dye degradation. As the contact area of the dielectric surface with the conducting liquid (cathode) increases, the dye degradation rate increases. We observed a nearly 20-25\% reduction in degradation time by changing the volume of DBD plasma in the gap. The concentration of reactive oxidative species is expected to increase with increasing plasma volume. The higher concentration of oxidative species gives a higher degradation rate of the dye in a given solution. These results helped in optimizing the contact area between the tube (dielectric) and the conducting water solution (cathode) to achieve maximum dye degradation efficiency. 
Therefore, the dielectric tube was fully immersed in the conducting solution (cathode) [see Fig.\ref{fig:fig1}] to achieve the maximum DBD plasma volume that favours the higher degradation rate. 
\subsection{Effect of initial TDS of solution on degradation rate}
In the textile industries, normal tap water or drinking water with a higher TDS value is used. But in most laboratory experiments, model wastewater solutions (dye solutions) are prepared in distilled or DI water.  
To examine the role of water used in preparing the dye solutions in the degradation mechanism in post-discharge configuration, two sets of experiments were performed using normal tap water (high TDS and EC) and distilled water (Low TDS and EC) dye solutions. The degradation times of different concentrations of dye solutions, prepared in normal water and distilled water, are presented in Table \ref {table:table1}. It is observed that approximately 10 to 15\% more time is required to degrade the same amount of crystal violet in normal water compared to distilled water. The presence of inorganic/organic molecules in tap water reduces the availability of oxidative species to present (added) dye molecules. It shows that the post-discharge degradation mechanism of dye does not have a significant impact on the initial TDS and EC of dye solutions. This important observation provides a futuristic path for using normal water in preparing model wastewater samples (for research purposes at the lab scale) to correlate with real textile industrial wastewater. 
\subsection{Role of the gas flow rate}
\begin{figure} 
    \centering
    \includegraphics[width=0.65\linewidth]{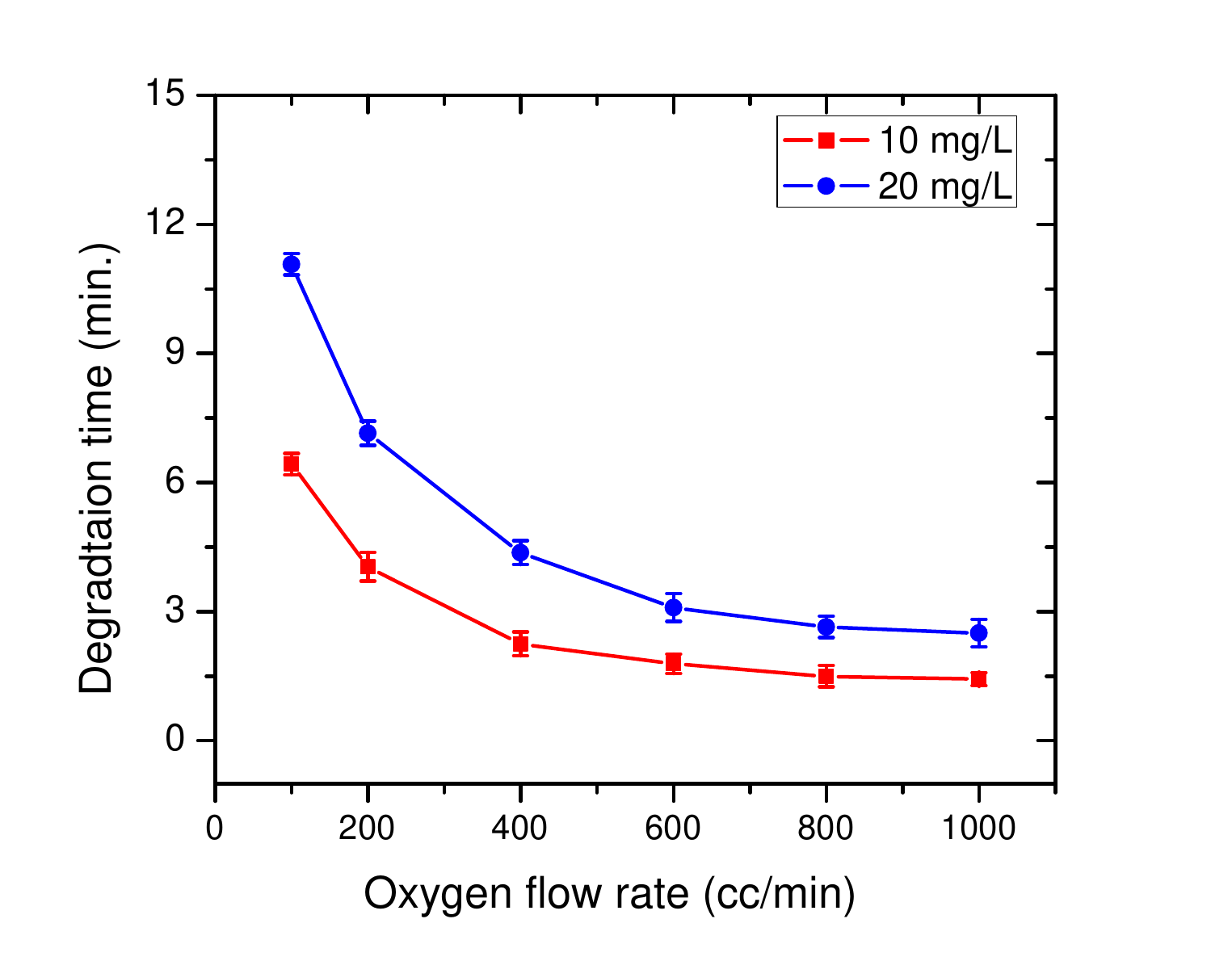},
    \caption{\textit{} Variation of degradation time (at which dye degraded $>$ 98 \%) with different oxygen flow rates at two different CV dye concentrations, 10 mg/L and 20 mg/L. The volume of dye solution, Magnetic stirrer speed and temperature of the solution were 500 ml, 400 RPM, and  26.4\textdegree C, respectively.}
    \label{fig:fig4}
\end{figure}

It is expected that the input oxygen flow rate may decide the mass transfer rate of reactive species to the dye solution in the post-discharge configuration. Therefore, a set of experiments with similar experimental conditions but different gas flow rates was conducted. The volume of the CV solution was 500 ml, the concentration of the solution was 20/10 mg/L, and the speed of the magnetic stirrer was 400 RPM in each experiment. The variation in dye degradation time with respect to oxygen flow rates is shown in Fig.\ref{fig:fig4}. We observed an exponential decrease in degradation time with increasing gas flow rate. A higher degradation rate is observed with increasing the gas flow rate up to 500 cc/min, and it falls slowly with further increasing the oxygen flow rate. Fig.\ref{fig:fig14} shows that the concentration of ozone in reactive gases increases linearly with oxygen flow rates. Although the dye degradation rate does not increase linearly above 500 cc/min gas flow rate. It was observed that higher gas flow rates produce larger gas bubbles. The larger the bubble size, the smaller the total surface area and bubble density; consequently, reactive gas-liquid mass transfer decreases. It may lead to a decrease in the equilibrium concentration of reactive oxidative species (mainly $O_3$) in the dye solution. Thus, the results suggest that an appropriate input gas flow rate can be set to achieve a higher degradation rate, considering other factors as well. 
\subsection{The effect of stirrer speed (RPM) on degradation}
\begin{figure}
    \centering
    \includegraphics[width=0.65\linewidth]{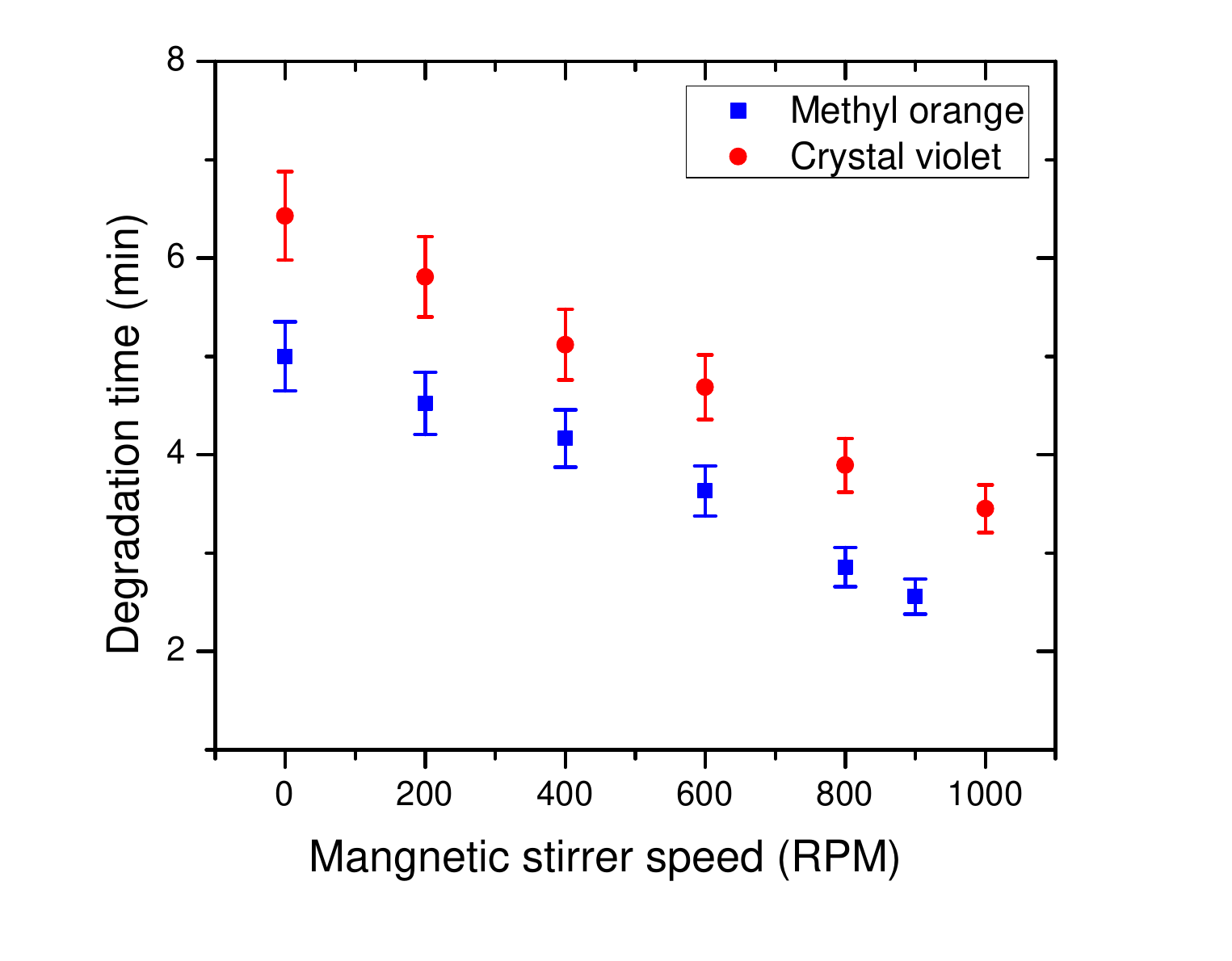},
    \caption{\textit{} Variation of degradation time for CV and MO dye solutions (at which dye degraded $>$ 98 \%) with different stirrer speeds (RPM). Volume of dye solution was 500 ml, dye concentration was 40 mg/L, gas flow rate was 400 cc/min, and the dye solution temperature was 28\textdegree C.}
    \label{fig:fig5}
\end{figure}

In the post-discharge configuration, reactive gaseous species (oxidative agents) are dissolved and diffused into the dye solution, leading to chemical reactions with the dye molecules/organic molecules. The rate of reaction could be directly proportional to the rate of mass transfer (diffusion in volume) of oxidative species in the solution. Is it possible to increase the diffusion rates of oxidizing species into the dye solution by stirring it externally? We investigated the role of stirring speed on the degradation time for a particular dye. The experiments were performed using two different dye solutions (crystal violet and methyl orange) at the same concentration (40 mg/L) and gas flow rate (400 cc/min). Fig.\ref{fig:fig5} depicts the change in degradation time with changing stirrer speeds. The degradation time reduces linearly with increasing the stirrer speeds in rotations per minute (RPM). The observations predict that the higher mixing or diffusion of reactive species (or oxidizing species) into the dye solution can be achieved by stirring the solution using external sources.     
\subsection{The effect of initial crystal violet concentration}
\begin{figure}
    \centering
    \includegraphics[width=0.65\linewidth]{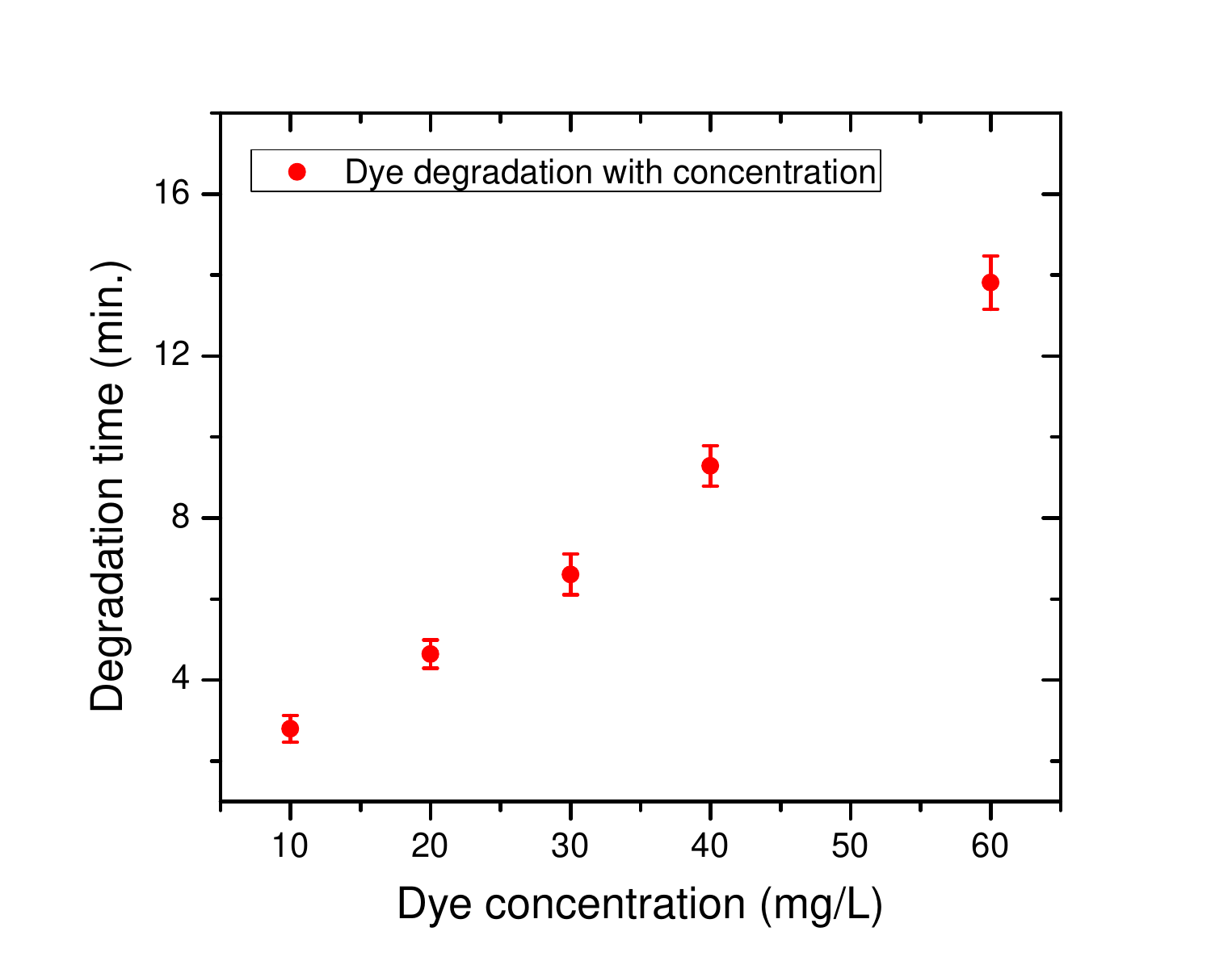},
    \caption{\textit{} Degradation time for CV (at which dye degraded $>$ 98 \%) with changing concentration of dye solution. The volume of dye solution, gas flow rate, stirrer speed, and the temperature of dye solution were 500 ml, 400 cc/min, 400 RPM, and 25\textdegree C, respectively.}
    \label{fig:fig6}
\end{figure}

A set of experiments is conducted to investigate the effect of initial dye concentration on the degradation rate at a given gas flow rate. The degradation time was recorded for different initial concentrations of dye solutions, while maintaining similar conditions throughout the experiments. The degradation time at different dye concentrations (10 mg/L to 60 mg/L) is displayed in Fig.\ref{fig:fig6}. We noticed a nearly linear relationship between degradation time and dye concentration. If the concentration of dye is doubled, approximately twice the time is required to achieve the complete degradation in the post-discharge configuration. These findings help in determining the optimal plasma operation time for low- or high-concentration dye solutions or wastewater.    
\subsection{Role of dye solution temperature}
\begin{figure}
    \centering
    \includegraphics[width=0.65\linewidth]{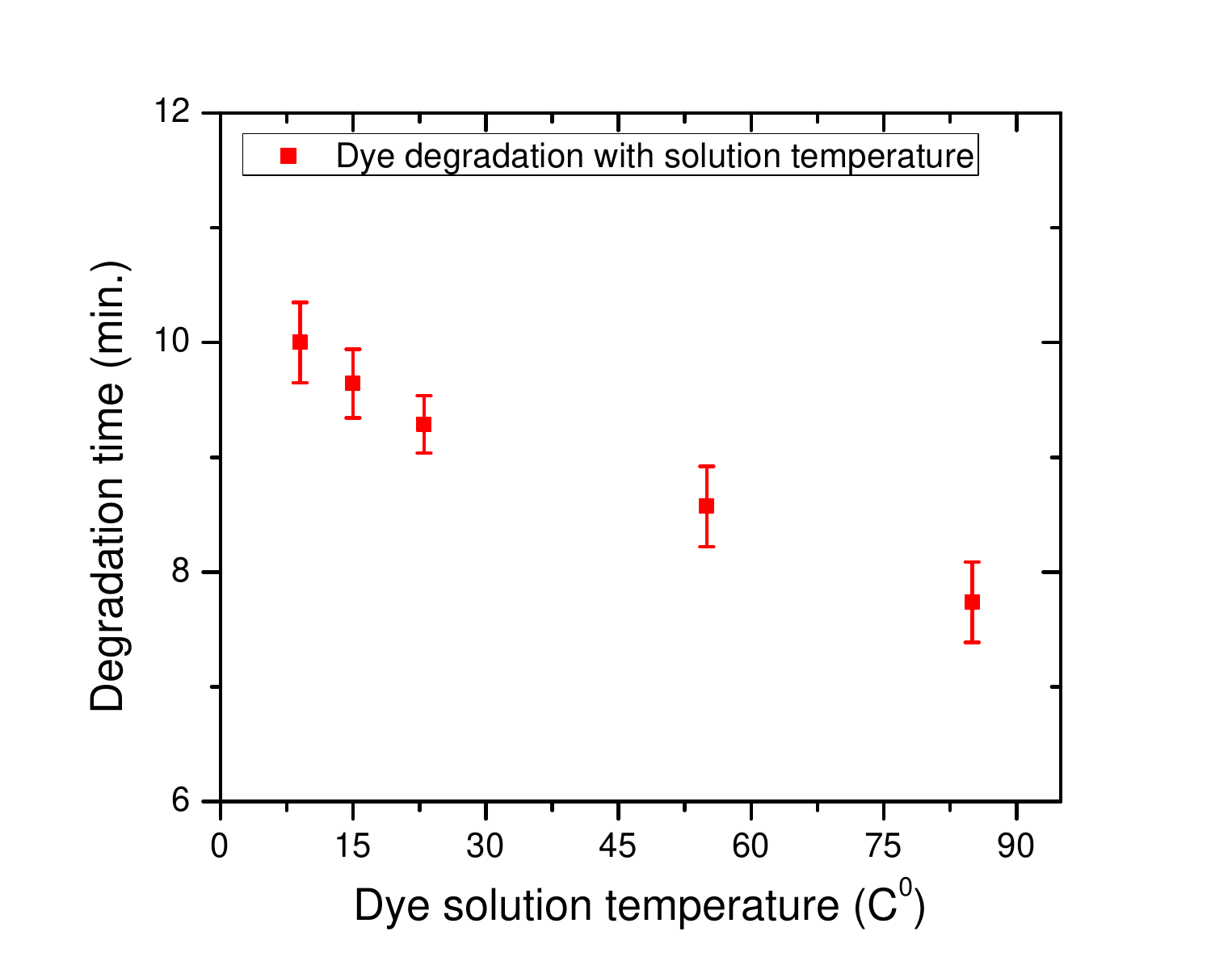},
    \caption{\textit{} Degradation time for CV (at which dye degraded $>$ 98 \%) with changing temperature of dye solution. Volume of dye solution was 500 ml, dye concentration was 40 mg/L, gas flow rate was 400 cc/min, stirrer speed was 400 RPM, and the temperature of dye solution was 28\textdegree C.}
    \label{fig:fig7}
\end{figure}

The impact of the temperature of the dye solution on the degradation rate in the post-discharge configuration is investigated by performing a set of experiments with similarly prepared dye solutions in cold and hot water. The variation of degradation time with the temperature of the dye solution is depicted in Fig.\ref{fig:fig7}. The dye degradation rate was observed to be higher in the cold dye solution as compared to the hot dye solution.
A nearly 25 to 30 \% decrease in degradation time was reported with changing the solution temperature from 80\textdegree C to 20\textdegree C. The solubility of oxidative species decreases with increasing solution temperature, resulting in a decrease in the mass transfer rate of reactive species from the gas phase to the liquid phase. In the meantime, the chemical reaction rate increases with the increasing kinetic energy of the reactants as the temperature rises. However, the dominant role of mass transfer rate over the chemical reaction rate with temperature reduces the dye degradation rate. The results could help in achieving higher degradation efficiency of complex molecules/dyes by controlling the temperature of the dye solution/wastewater. 
\subsection{The effect of pH of dye solution on degradation}
\begin{figure}
    \centering
    \includegraphics[width=0.65\linewidth]{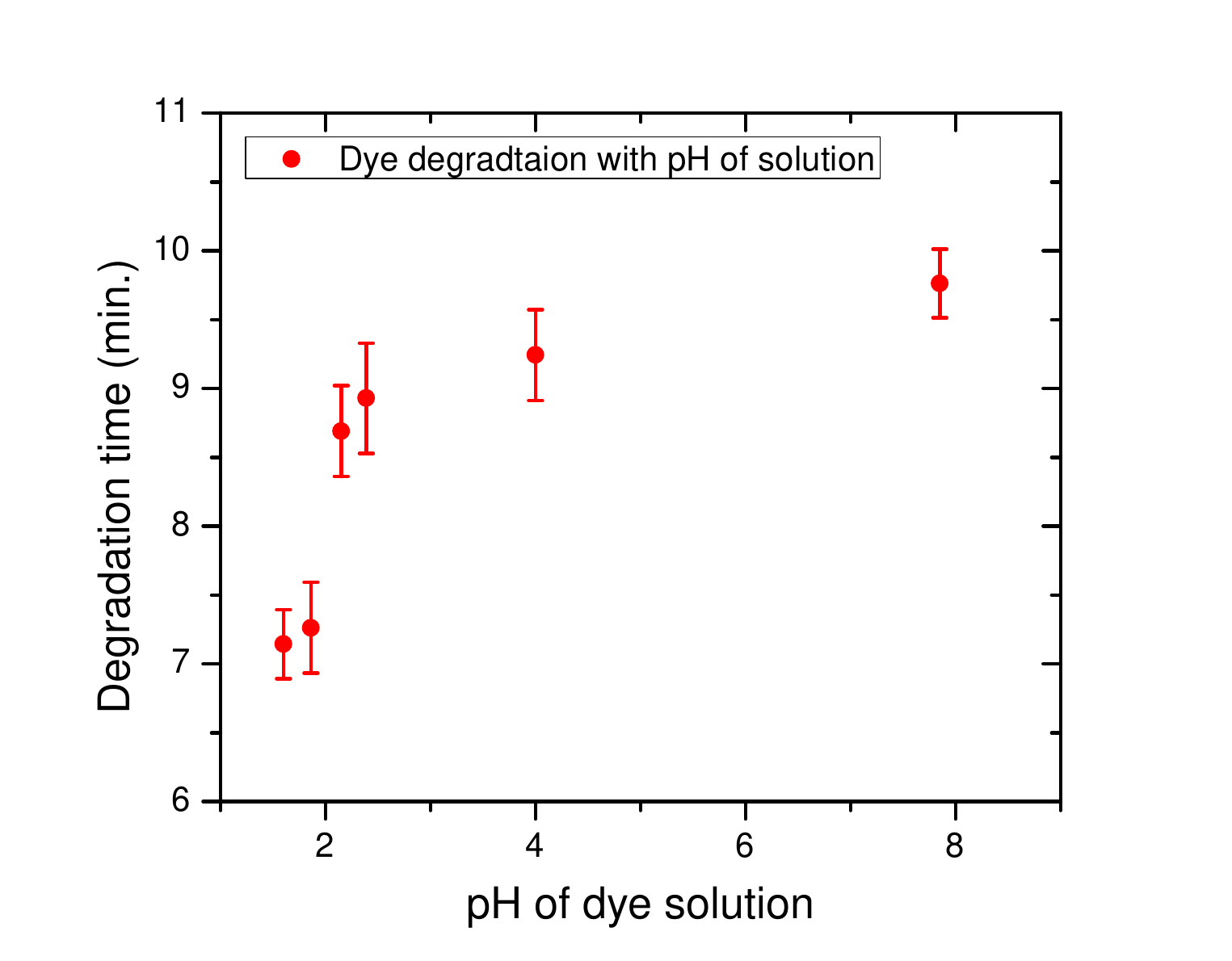},
    \caption{\textit{} Degradation time (at which dye degraded $>$ 98 \%) with changing pH of crystal violet dye solution. Volume of dye solution was 500 ml, gas flow rate was 400 cc/min, stirrer speed was 400 RPM, concentration of dye was 40 mg/L, and the temperature of the solution was 25.2\textdegree C.}
    \label{fig:fig8}
\end{figure}

It is a fact that the pH of the solution may play a significant role in controlling the chemical reactions between oxidizing agents and organic complex molecules. To investigate the effect of the dye solution's pH on its degradation rate, several experiments were conducted using dye solutions of varying pH levels. The pH of the solution was changed by adding sulfuric acid to the prepared dye solution of given volume and concentration. In an acidic environment (low pH) of the solution, the rate of degradation is found to be higher as compared to the mild basic nature of the crystal violet solution (see Fig.\ref{fig:fig8}). We reported a nearly 20 to 25\%  increase in CV degradation time with increasing the pH of the solution from 2 to 8. The positive impact of pH on dye removal can be attributed to the presence of strong oxidative species ($\dot{OH}$) in the solution. The acidic nature of the dye solution enhances the oxidation rate of organic (dye) molecules in the post-discharge configuration.   
    \subsection{Effect of conductivity of dye solution}
\begin{figure}
    \centering
    \includegraphics[width=0.65\linewidth]{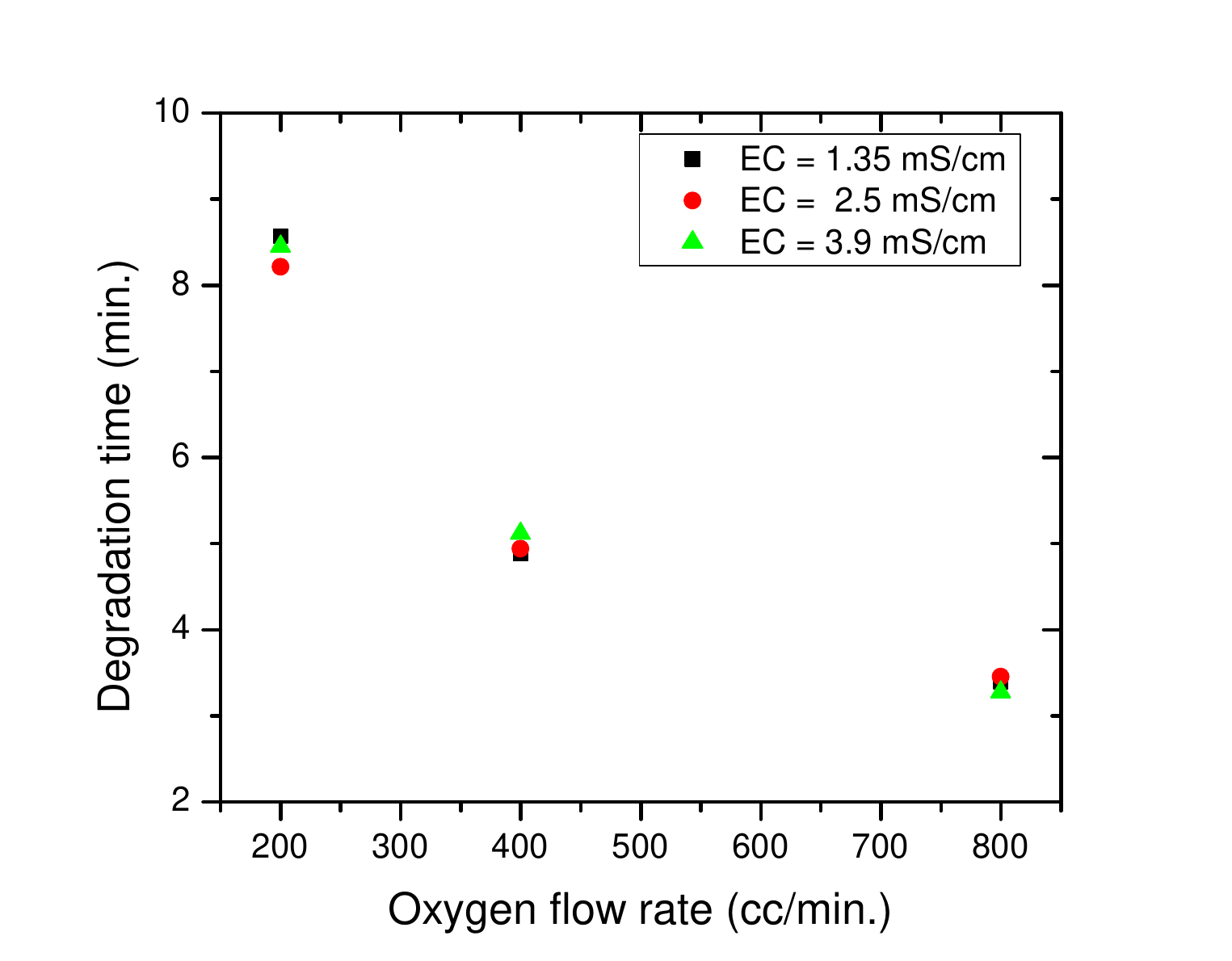},
    \caption{\textit{} Degradation time for CV (at which dye degraded $>$ 98 \%) with changing electrical conductivity of dye solution. Volume of dye solution, concentration of dye, gas flow rate, speed of stirrer and the temperature of dye solution were 500 ml, 40 mg/L, 400 cc/min., 400 RPM and 28\textdegree C, respectively.}
    \label{fig:fig9}
\end{figure}

In direct plasma-dye solution interaction \cite{vyas2023degradationmethylenebluedye}, the electrical conductivity plays a significant role in the degradation rate of the dye. To investigate the effect of electrical conductivity on the degradation rate of the dye solution, several experiments were conducted using dye solutions of varying conductivities, treated at different gas flow rates. In Fig.\ref{fig:fig9}, the data points of three experiments (different EC) at a given flow rate coincide, indicating the negligible role of the conductivity of the crystal violet dye solution on the degradation rate. It demonstrates that oxidation processes in post-discharge DBD configuration are independent of the presence of charged particles/other species in dye solutions.  
\section{Kinetics of crystal violet degradation} \label{sec:sec5}
\begin{figure}
    \centering
    \includegraphics[width=0.65\linewidth]{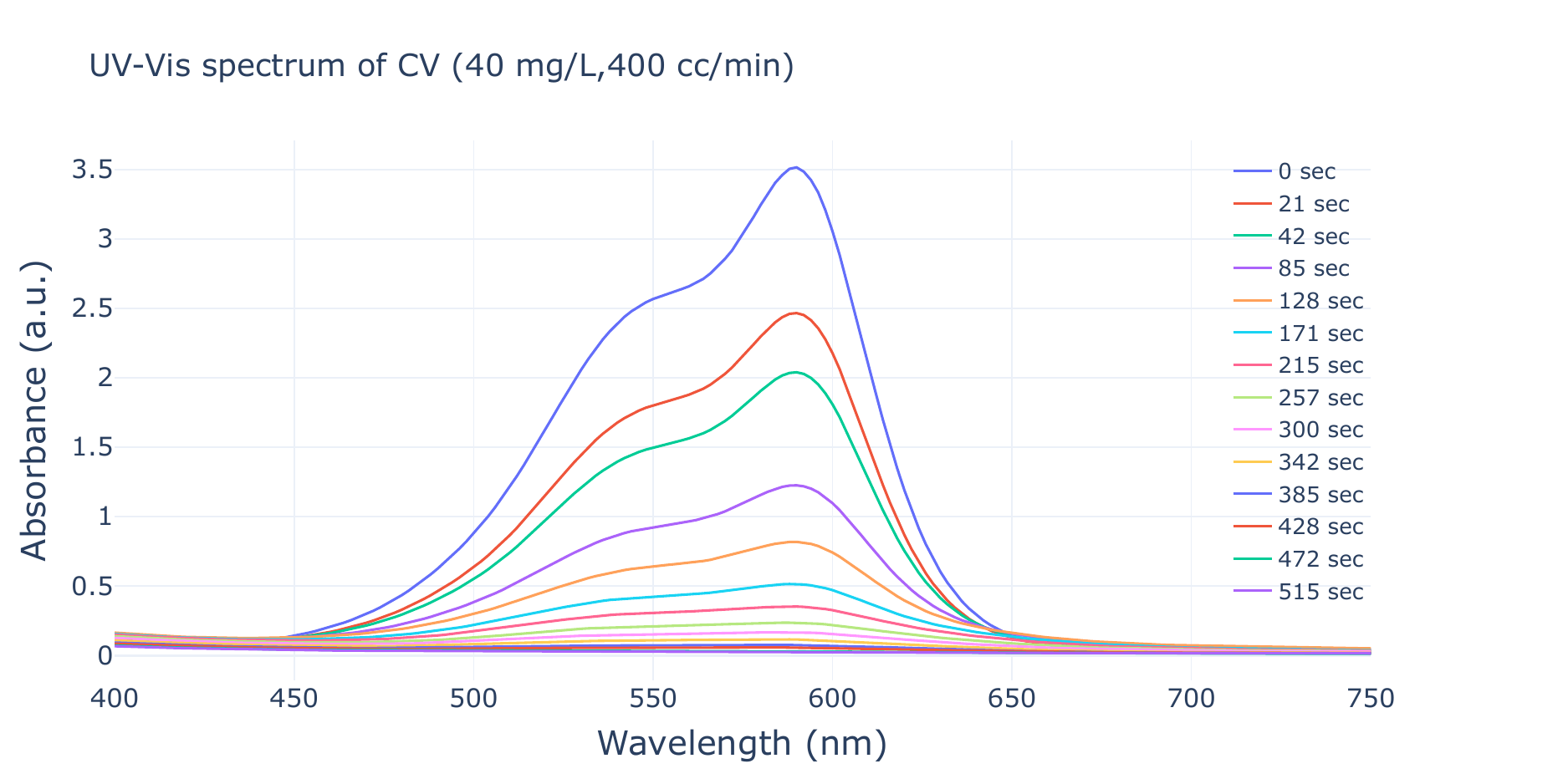},
    \caption{\textit{} UV-Vis absorption spectrum of crystal violet at different treatment times. Volume of dye solution was 500 ml, dye concentration was 40 mg/L, gas flow rate was 400 cc/min, stirrer speed was 400 RPM, and the temperature of the dye solution was 27.6\textdegree C.}
    \label{fig:fig10}
\end{figure}

\begin{figure}
    \centering
    \includegraphics[width=0.65\linewidth]{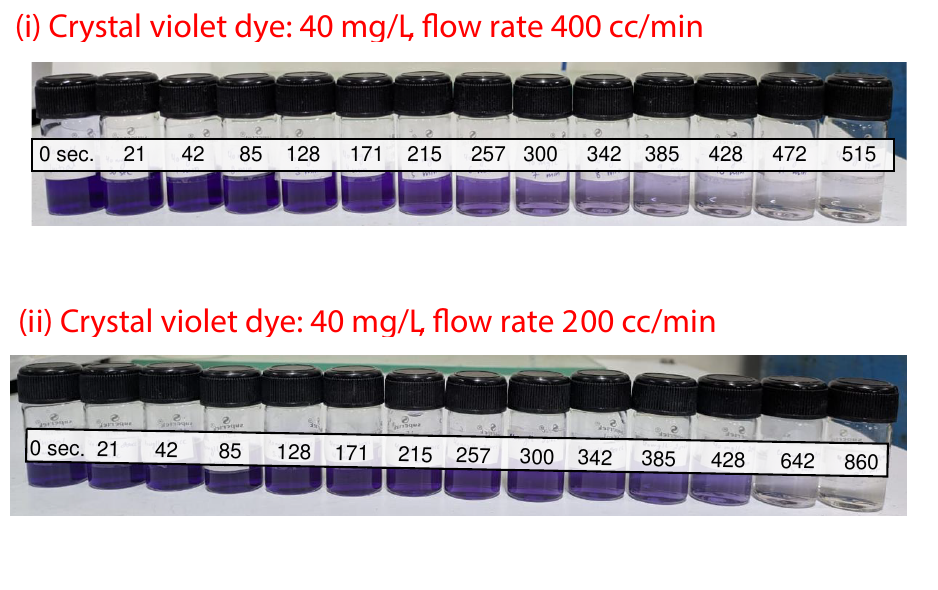},
    \caption{\textit{}{Change in colour of crystal violet (CV) dye solution of concentrations 40 mg/L at gas flow rates of 400 cc/min and 200 cc/min. (i) for 40 mg/L CV solution, at 400 cc/in, $t_0$ = 0 Sec. and $t_{max}$ = 515 Sec. (8.5 min), (ii) for 40 mg/L solution at 200 cc/min, $t_0$ = 0 Sec. and $t_{max}$ = 860 Sec. ($\sim$ 20 min). The dye solution becomes transparent, like normal water, once more than 96\% of the dye molecules have been decomposed.}}
    \label{fig:fig11}
\end{figure}

\begin{figure*} 
 \centering
\subfloat{{\includegraphics[scale=0.180050]{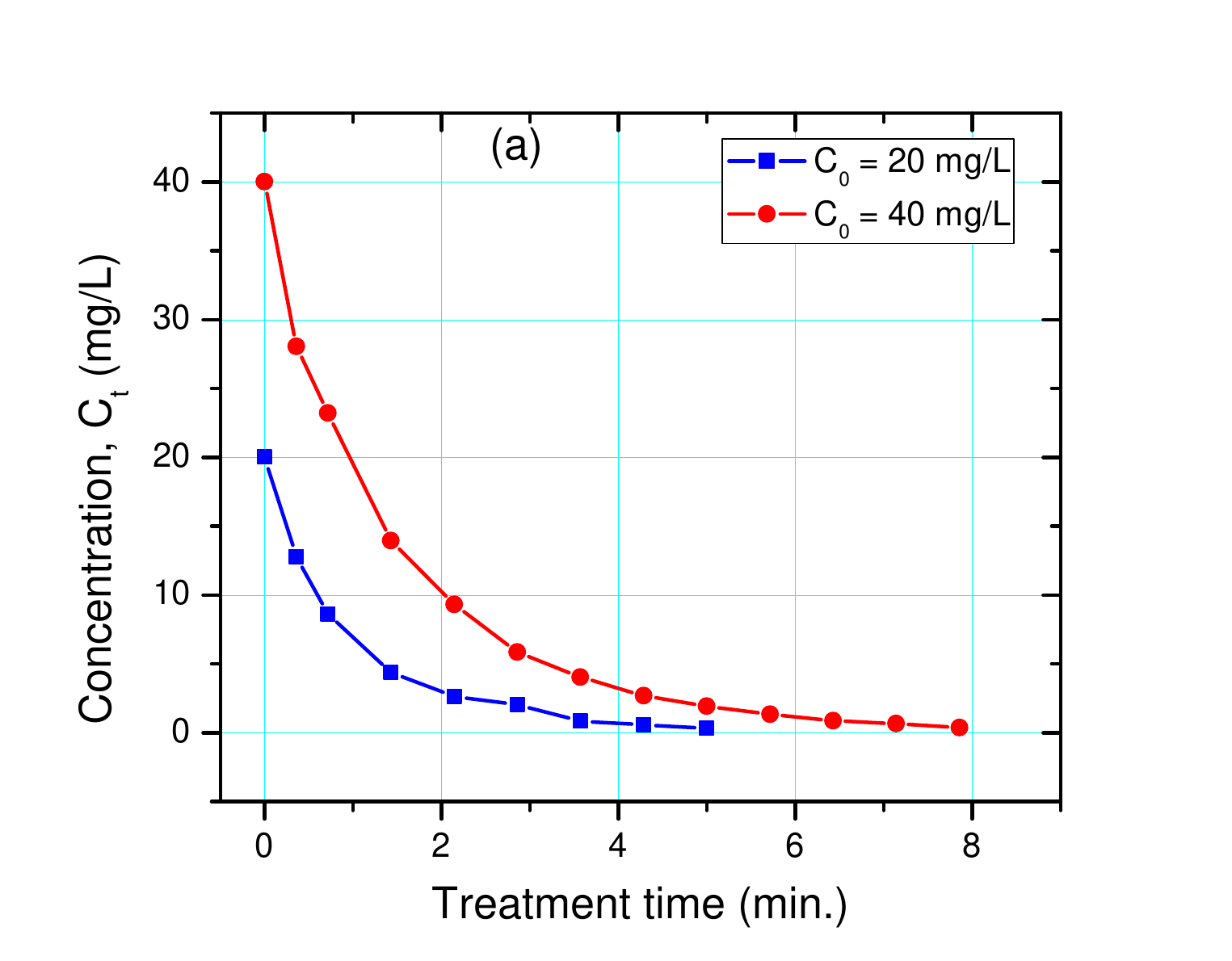}}}%
 \subfloat{{\includegraphics[scale=0.18050]{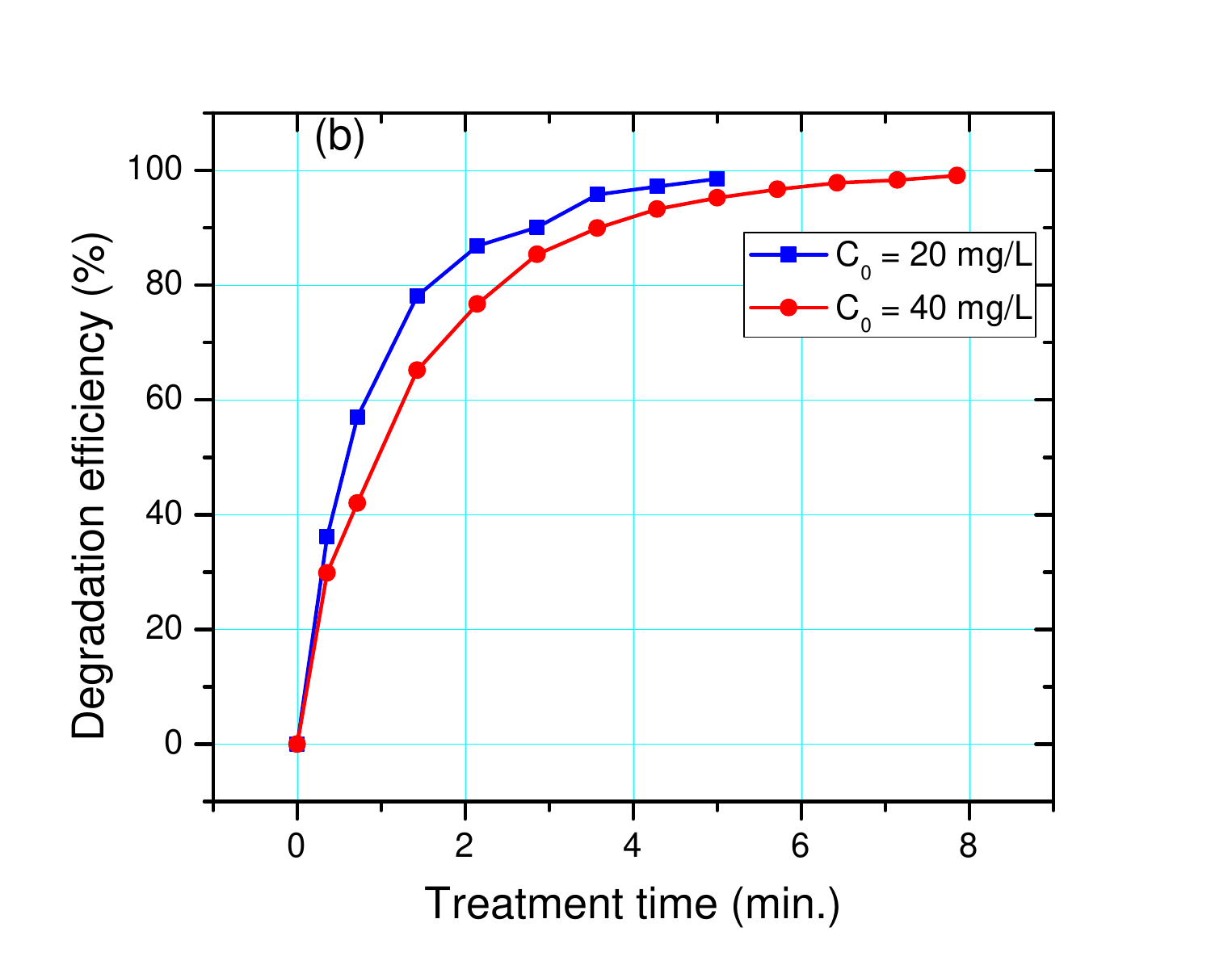}}}
\quad
\subfloat{{\includegraphics[scale=0.20050]{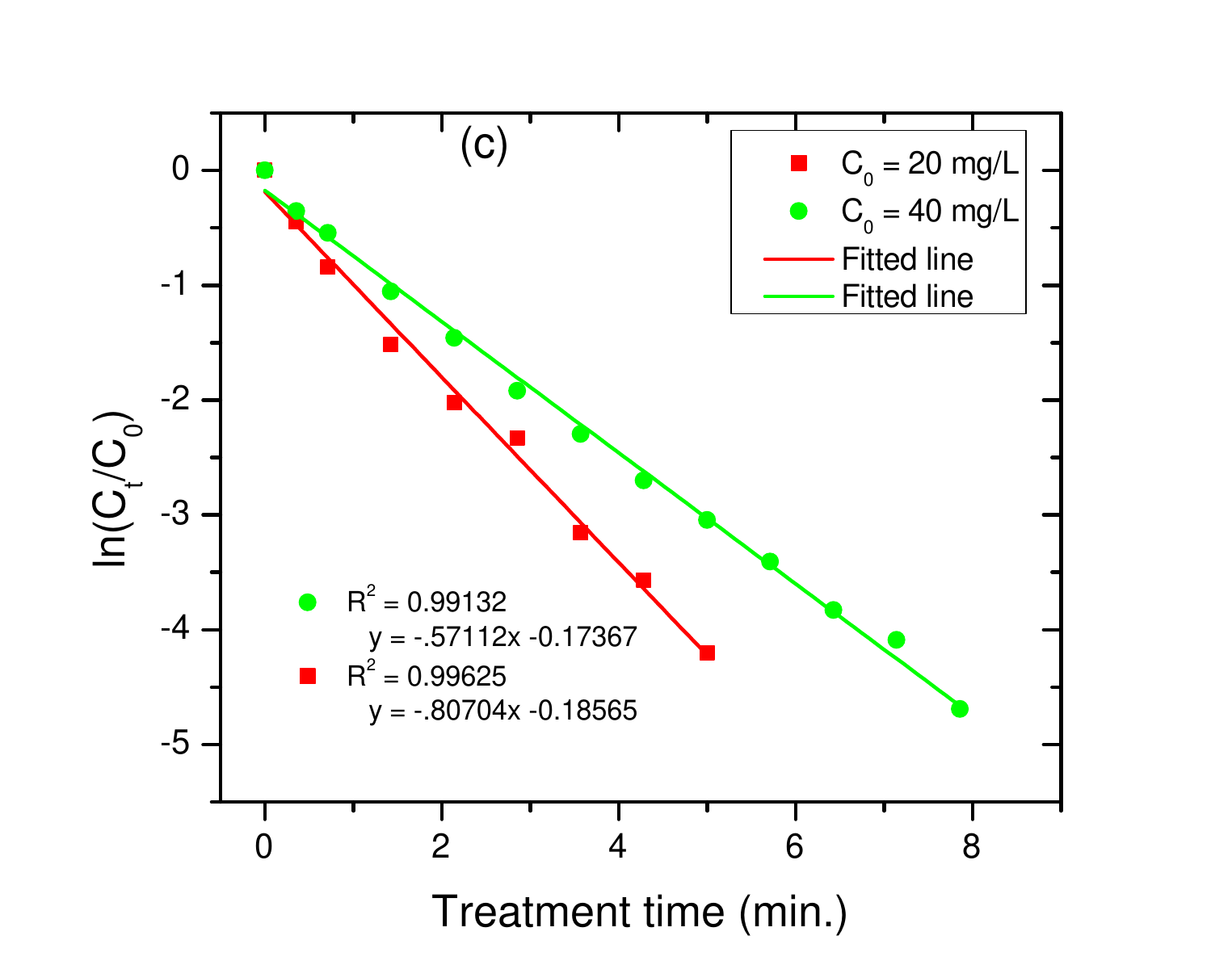}}}
\caption{\label{fig:fig12} (a) CV concentration variation with treatment time. (b) crystal violet degradation efficiency(\%) with treatment time. (c) reaction kinetics for 20 mg/L and 40 mg/L at a gas flow rate of 400 cc/min. The volume and temperature of the solution were 500 ml and 26.5\textdegree C, respectively.} 
\end{figure*}

In order to determine the order of the degradation process and the specific rate constant, kinetic study of crystal violet degradation was performed for two different concentrations (20 mg/L and 40 mg/L) at a constant flow rate (400 cc/min.) and two different gas flow rates (200 cc/min. and 400 cc/min.) at a given concentration (40 mg/L) of CV. The CV samples of 5 ml were collected at different times during the plasma treatment for a particular experiment. The collected samples were further analyzed using the UV-Vis absorption spectrometer. The UV-Vis absorption spectrum of CV of 40 mg/L at 400 cc/min is displayed in Fig.\ref{fig:fig10}. The maximum absorption peak appears at a wavelength of 590 nm, corresponding to crystal violet (from an extended chromophore), which represents the concentration of the dye in the water solution. The maximum absorption peak at 590 nm corresponds to crystal violet before the treatment. Lowering the absorption peaks at 590 nm or slightly lower value (Fig.\ref{fig:fig11}) with plasma treatment is due to the fragmentation of the aromatic ring links by oxidation processes.
The changes in color of the CV dye solution with treatment for two different gas flow rates are shown in Fig.\ref{fig:fig11}. It is possible to correlate the changes in colour of 40 mg/L CV solution (Fig.\ref{fig:fig11}) with the absorption peak values (Fig.\ref{fig:fig10}) at different treatment times.\\ 
The peak values at 590 nm in the absorption spectrum of CV were recorded to determine the variation of CV concentration with plasma treatment time under given experimental conditions. The variation of CV concentration ($C_t$) with treatment time at an oxygen flow rate of 400 cc/min is presented in Fig.\ref{fig:fig12}(a). The dye degradation efficiency is determined by the following equation:
\begin{equation}
Efficiency ~(\%) = \bigg(\frac{C_0 -C_t}{C_0} \bigg) \times 100
\end{equation}
where $C_0$ is an initial dye concentration and $C_t$ is the dye concentration at a given time t. 
The degradation efficiency of different dye concentrations is shown in Fig.\ref{fig:fig12}(b). It is clear from Fig.\ref{fig:fig12} that more than 90\% dye degradation takes place in 4 minutes of treatment, and the remaining 10\% degradation needs more than 4 minutes. It is possible to achieve more than 98\% of dye degradation in a few minutes (6 to 7 min.) of plasma treatment. In Fig.\ref{fig:fig12}(a), the plots of ${C_t}$ against treatment time suggest a first-order kinetic, which is described by the following rate equation (reference):
\begin{equation} \label{eq:eq1}
   -\frac{dC_t}{dt} = k \times C_t
\end{equation}
where, $C_t$ is the CV concentration at time t, $k$ represents the first-order rate constant. 
After integration of Eq.\ref{eq:eq1} with initial conditions, the first-order rate equation becomes:
\begin{equation}
   ln \bigg(\frac{C_t}{C_0}\bigg) = -k \times t
\end{equation}
where, $C_0$ is the initial CV concentration. 
This equation was used to verify the order of plasma degradation process and rate constant from the CV concentration data (Fig.\ref{fig:fig12}(a)).
It is clearly seen in Fig.\ref{fig:fig12}(c) that $ln(\frac{C_t}{C_0})$ against treatment time gives a straight line. This indicates that the oxidative degradation reactions follow pseudo-first-order kinetics for both concentrations. The observed first-order rate constants and correlation coefficient ($R^2$) for CV of 20 mg/L and 40 mg/L at a gas flow rate of 400 cc/min are 0.57112 $min.^{-1}$, 0.99132 and 0.80704 $min.^{-1}$, 0.99625 respectively. 
\begin{figure*} 
 \centering
\subfloat{{\includegraphics[scale=0.180050]{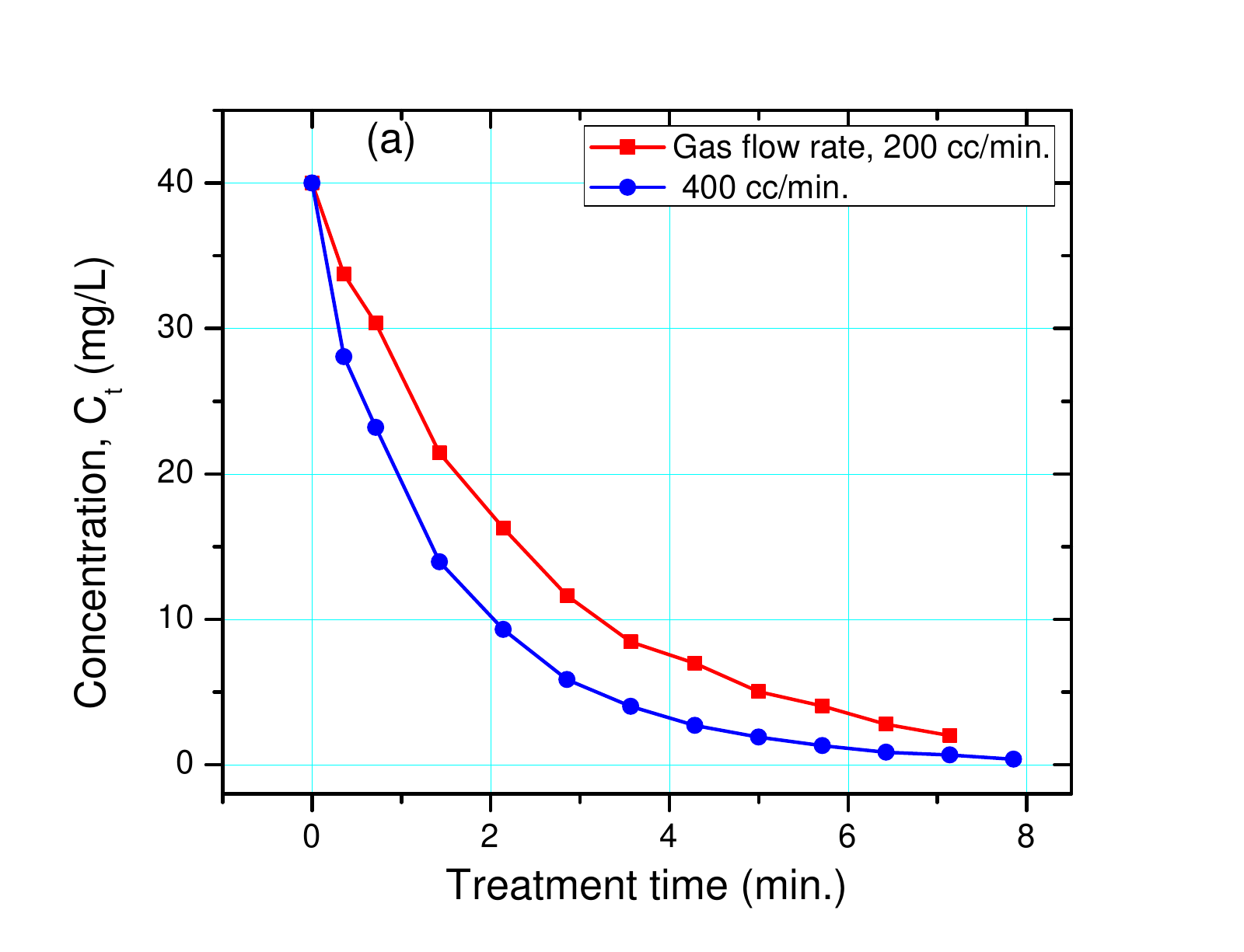}}}%
 \subfloat{{\includegraphics[scale=0.1805]{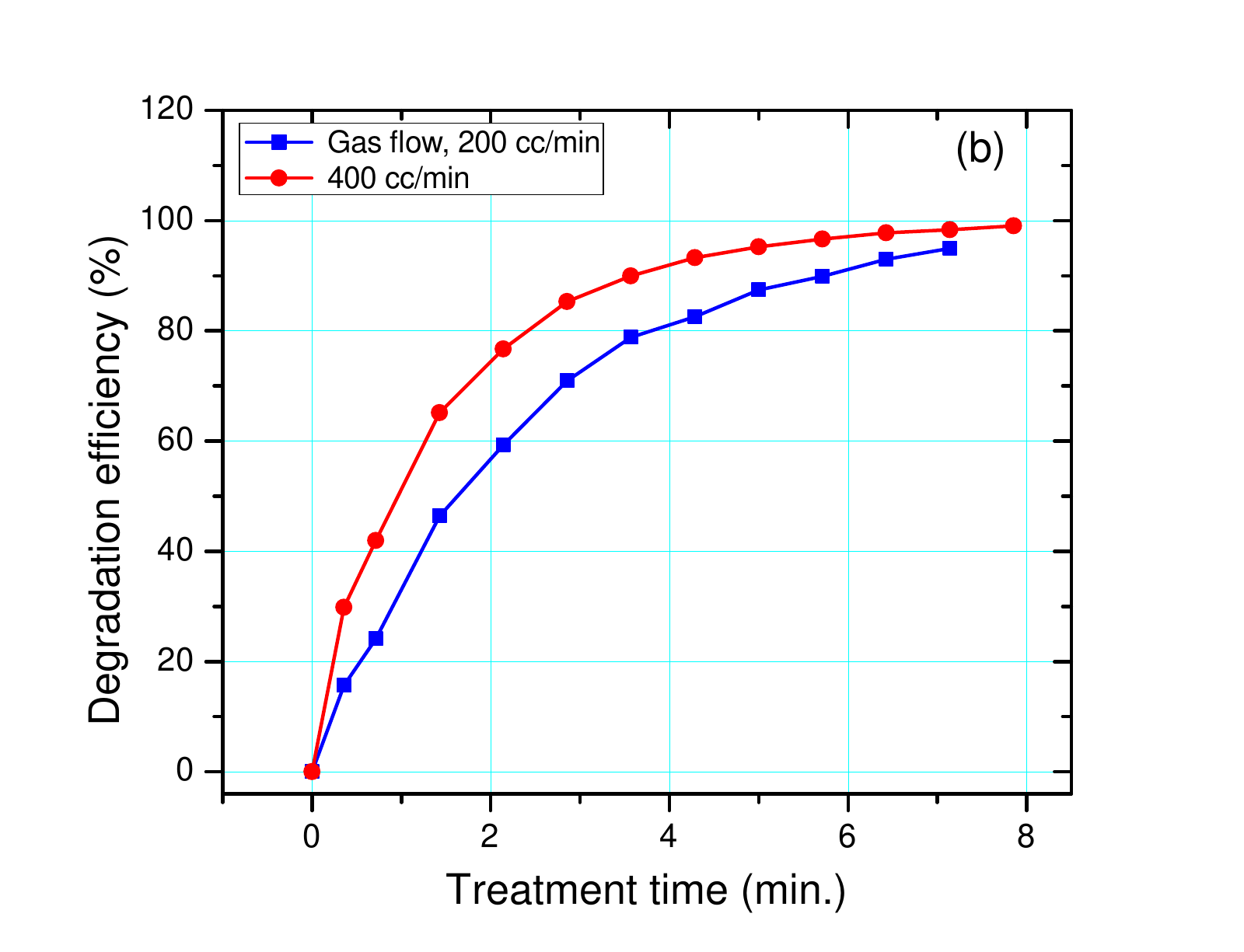}}}
\quad
\subfloat{{\includegraphics[scale=0.20050]{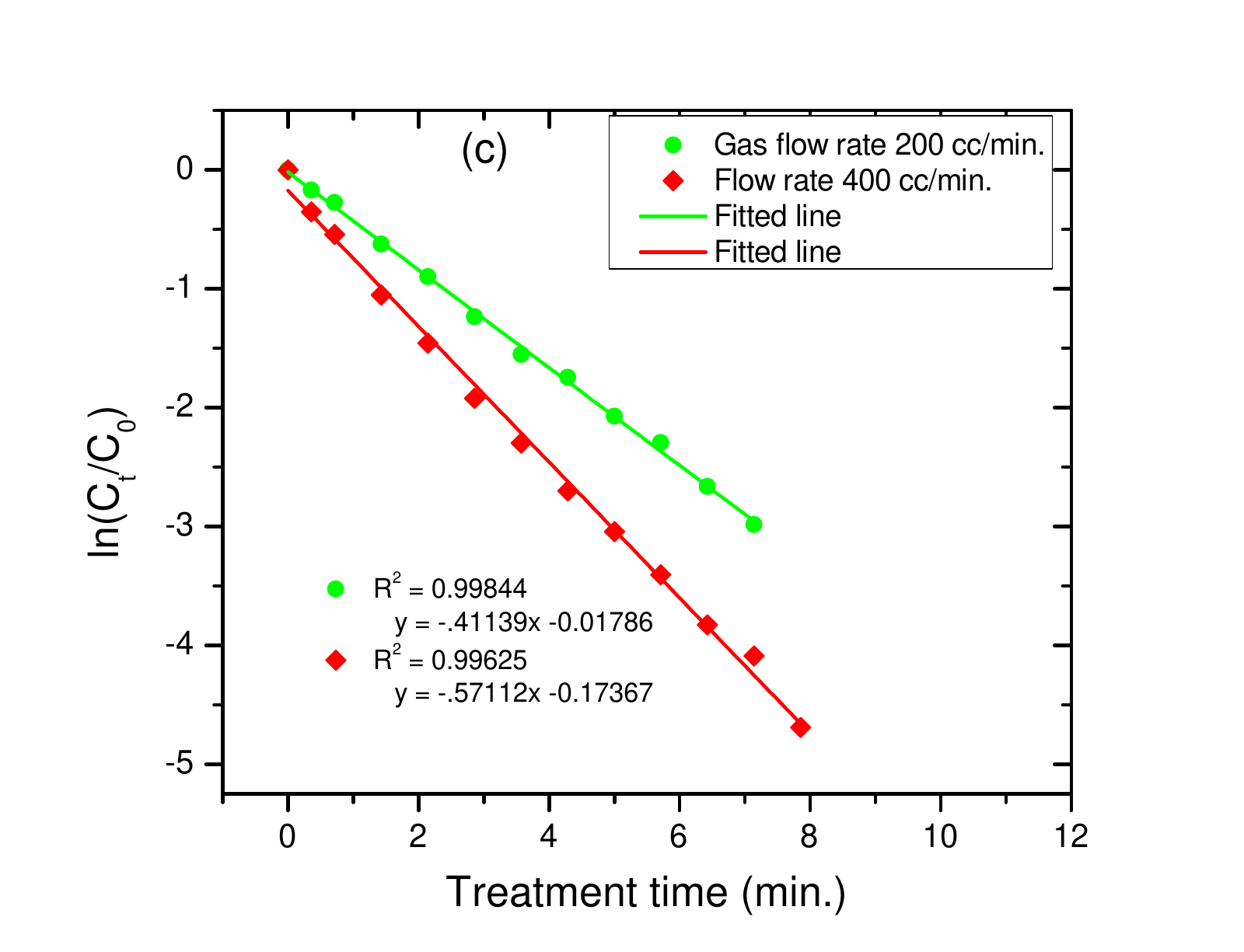}}}
\caption{\label{fig:fig13} (a) CV concentration variation with treatment time. (b) crystal violet degradation efficiency (\%) with treatment time. (c) reaction kinetics for 40 mg/L at gas flow rates of 200 cc/min and 400 cc/min. The volume and temperature of the solution are 500 ml and 26.5\textdegree C, respectively.} 
\end{figure*}
\\
In Fig.\ref{fig:fig13}(a), the variation of CV concentrations when 40 mg/L solutions were treated with gas flow rates of 200 cc/min and 400 cc/min while keeping other experimental conditions similar. The efficiency of CV degradation with two different gas flow rates (200 cc/min and 400 cc/min) is depicted in Fig.\ref{fig:fig13}(b). It is clear from Fig.\ref{fig:fig13}(b) that more than 80\% dye degradation takes place in 4 minutes of treatment, and the remaining 20\%  degradation requires more than 4 minutes. The CV concentration data (Fig.\ref{fig:fig13}(a)) were used to verify the order of the degradation reaction.    
The plots between $ln(\frac{C_t}{C_0})$ and treatment time at 200 cc/min and 400 cc/min gas flow rates are shown in Fig.\ref{fig:fig13}(c). Fig.\ref{fig:fig13}(c) indicates that the oxidative degradation reactions follow pseudo-first-order kinetics for both gas flow rates. The observed first-order rate constants and correlation coefficients ($R^2$) for CV of 40 mg/L at gas flow rates of 200 cc/min and 400 cc/min are 0.41139 $min.^{-1}$, 0.99844 and 0.57112 $min.^{-1}$, 0.99625 respectively.   
\section{Discussion} \label{sec:sec6}
\begin{figure}
    \centering
    \includegraphics[width=0.65\linewidth]{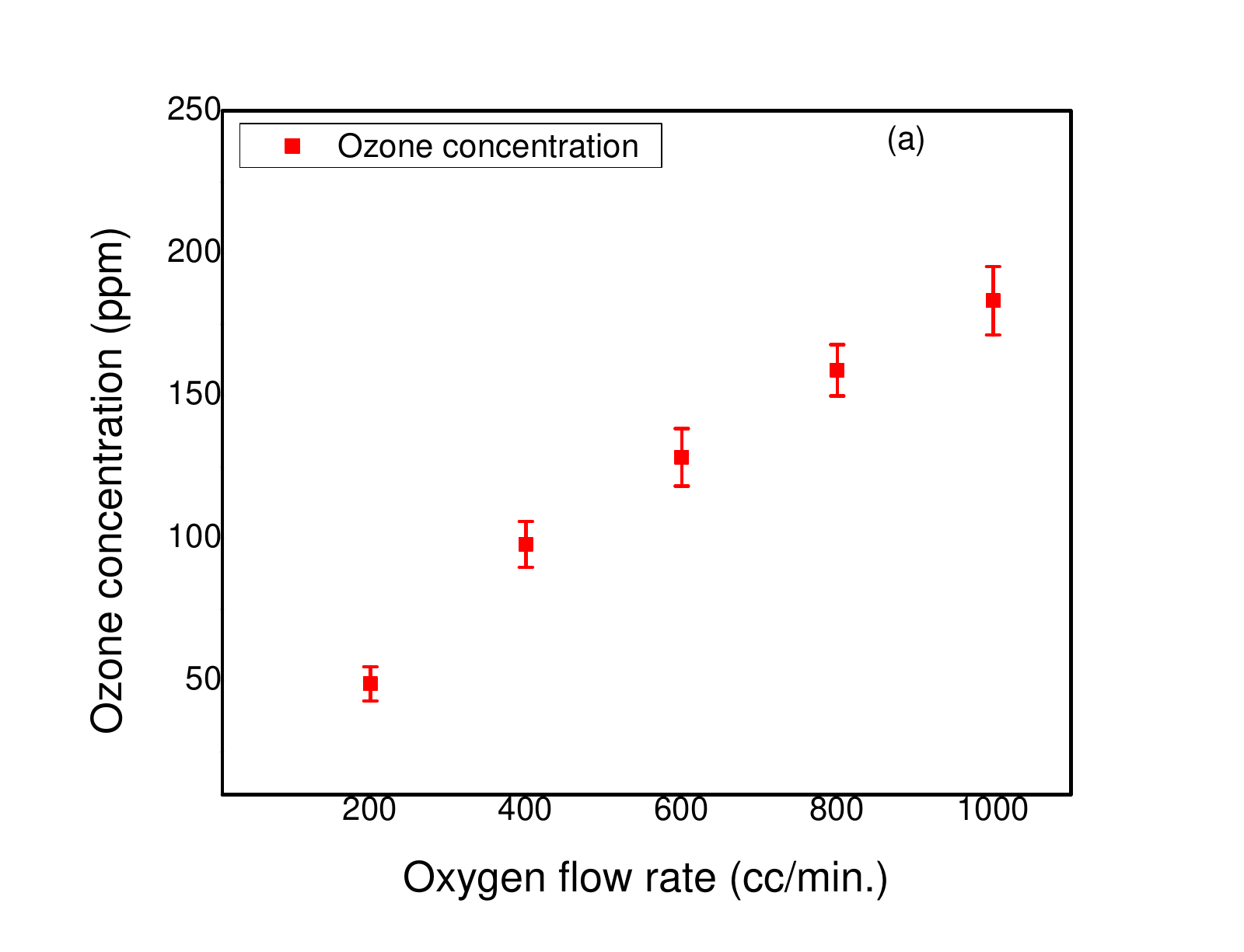},
    \caption{\textit{} Ozone concentration in post-discharge reactive gases with input oxygen flow rates to DBD reactor.}
    \label{fig:fig14}
\end{figure}

In the dielectric barrier discharge reactor, oxygen with 99\%  purity was used as input gas. Therefore, ozone is expected to be the major reactive species along with other long-lived species (Hydrogen Peroxide (\(H_{2}O_{2}\)), Hydroxyl Radical ($\dot{OH}$), Nitrogen Dioxide ($\dot{NO_{2}}$) etc.) in the post-discharge configuration. The reaction path for ozone creation in the discharge zone can be understood by the Eq.\ref{eq:eq4}
\begin{equation} \label{eq:eq4}
    e^{-} + O_2 \rightarrow 2O + e^{-}
    O + 2O_2 \rightarrow O_3 + O_2
\end{equation}
By keeping this in mind, we measured the ozone concentration (in reactive gases) dissolved into the water at different input oxygen flow rates. The Iodometric titration technique was used to measure the concentration of ozone in the gaseous phase of reactive species emanating from the DBD reactor. 
A 10 mM potassium iodide (KI) solution was exposed to reactive gases emanating from the DBD reactor. The colour of the KI solution changed from clear to brownish yellow due to the formation of $I_2$ by the following oxidation reaction.

\begin{equation} \label{eq:eq5}
    O_3 + 2I^{-} + H_2O \rightarrow I_2 + 2OH^{-}+O_2
\end{equation}
As per the reaction, the number of moles of $O_3$ is proportional to the number of moles of $KI$. To measure the moles of $I_2$, the solution was titrated using a 0.05 M sodium thiosulfate solution, along with a 1 M acidic solution. The titration was continued until the solution's colour changed from brownish yellow to clear. 

\begin{equation}   \label{eq:eq6}
    I_2 + 2Na_2S_2O_3 \rightarrow 2NaI + Na_2S_4O_6
\end{equation}
 From both reactions (Eq.\ref{eq:eq5}\&\ref{eq:eq6}), it is clear that the number of moles of ozone is equal to half of the number of moles of sodium thiosulfate. The concentration of ozone can be estimated by knowing the titration parameters.
Fig.\ref{fig:fig14} illustrates the variation in ozone concentration at different oxygen flow rates, while maintaining constant discharge voltage and current. The results show that the ozone concentration increases with increasing gas flow rate up to 1 L/min. There may be a saturation region for ozone concentration at higher oxygen flow rates, but this was not a part of the present study. It indicates that the probability of electrons colliding with oxygen molecules increases with increasing gas flow rate, thereby increasing the rate of ozone formation. 
In the present post-discharge DBD configuration, ozone is expected to be the dominant oxidizing agent for the dye degradation processes (see Fig.\ref{fig:fig14}). The O$_3$ is dissolved into the dye solution and diffused to the whole volume by external stirring. The dissolved ozone can oxidize the dye molecules or any other complex organic molecules in different ways. The first one is a direct attack by ozone on the organic (dye) molecules. Due to the strong oxidative nature of ozone (oxidation potential is 2.07 V), it can easily break down complex dye structures, which leads to the formation of colourless fragments \cite{degradationcv_path3,degradtaion_pollutatnts_review_2022,ozonation_dye-degradation_path2}. The second possibility is to form a much stronger oxidative agent, hydroxyl radical $\dot{OH}$ (oxidation potential is 2.8 V) from dissolved $O_3$ in dye solution. The dissolved $O_3$ reacts with water or hydroxyl ions ($OH^{-}$) to form $\dot{OH}$. There is also a possibility to form $\dot{OH}$ from dissolved hydrogen peroxide ($H_2O_2$) in the dye solution. The attack of $\dot{OH}$ radicals on dye molecules breaks down its complex structure and cleavages aromatic rings and azo bonds. In direct or indirect oxidation processes, various smaller-sized fragments of complex dye molecules are formed \cite{degradationcv_path3,ozonation_dye-degradation_path2,CV_degrdation_path1}. Measuring the degraded fragments and analysing them were beyond the scope of the present work due to certain limitations.
\section{Validation of DBD reactor with other dyes} \label{sec:sec7}
\begin{figure}
    \centering
    \includegraphics[width=0.65\linewidth]{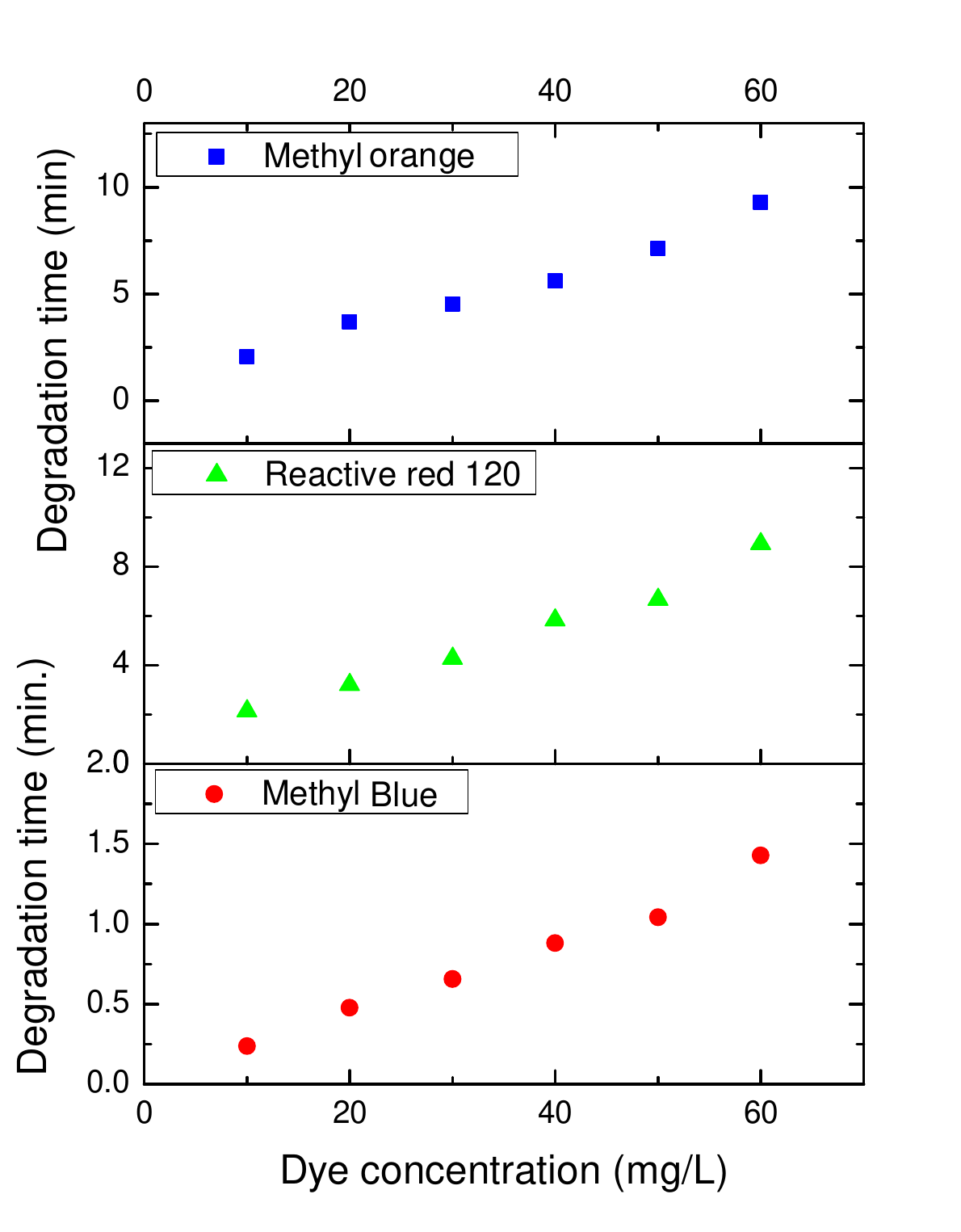},
    \caption{\textit{} Degradation time for different dyes (at which dye degraded $>$ 98 \%) with changing concentration of dye solution. Volume of dye solution was 500 ml, gas flow rate was 400 cc/min, stirrer speed was 400 RPM, and the temperature of the solution was 26.4\textdegree C.}
    \label{fig:fig15}
\end{figure}

After optimizing the DBD post-discharge reactor using crystal violet dye, a set of experiments is performed with three different dyes (Methyl orange, Reactive red 120, and Methyl Blue) to validate the applications of the newly designed and built DBD reactor in dye degradation (wastewater treatment). We have observed more than 98 \% degradations of complex dye molecules (at different concentrations) in a few minutes of post-discharge treatment (see Fig.\ref{fig:fig15}). Further studies will be conducted to understand the dye degradation pathways, reaction kinetics, and the formation of degraded fragments, as well as the impact of other experimental parameters. Detailed studies for these dyes (Methyl orange, Reactive red 120, and Methyl Blue) are a subject matter for future studies.         
\section{Summary and Future Outlook} \label{sec:sec8}
The work of this project is summarized as follows:
\begin{itemize}
    \item A newly post-discharge DBD reactor using a low-frequency (50 Hz) transient voltage source is designed and built for degrading complex organic pollutants at the laboratory scale.
    \item Crystal violet dye solutions were considered model textile industrial wastewater samples to validate the newly built DBD plasma reactor with liquid cathode (ground). 
    \item The operating conditions were validated using various experimental and wastewater (dye solution) parameters, including concentrations of dye, gas flow rate, EC/TDS of the solution, pH of the solution, temperature of the solution, EC of the cathode liquid, and volume of plasma, among others. 
    \item In post-discharge DBD configuration, the change to pH, EC and TDS of dye solutions after treatment is insignificant (or $<$ 2 to 4 \%).
    \item UV-Vis absorption spectroscopic technique was used to explore the reaction kinetics of CV degradation, and found that the oxidative degradation reactions follow pseudo-first-order kinetics.
    \item The Iodometric titration technique was used to measure ozone concentration in reactive gases. Therefore, ozone is considered a strong oxidant that plays a dominant role (direct or indirect) in degrading the complex dye structures and their fragments through oxidation. 
\end{itemize}
This primary study provides a futuristic path to upgrade the DBD reactor design for achieving the goal of mineralization of dye-containing wastewater. There is a need to degrade the larger-sized fragments of degraded dyes by integrating the oxidation process with other possible techniques to make it more reliable in direct applications. The appropriate research work in this line is continuing and will be communicated in future. 
\section{Acknowledgement} This work is supported by the Science \& Engineering Research Board (SERB) [Note: SERB has recently been subsumed by the Anusandhan National Research Foundation (ANRF)], a statutory body of the Department of Science \& Technology (DST), Government of India, under a Start-up Research Grant (SRG) with the sanction number [SRG/2023/000757]. The authors sincerely thank the Department of Physics and Astrophysics for allowing us to use the Central Experimental Facilities (CEF) for UV-Vis spectroscopy. The authors also thank Mr Vikram (Electronic lab) and Prof. Senthil Kumar for their technical support in understanding spectroscopic techniques. The authors are also grateful to the physics workshop at the Department of Physics and Astrophysics, University of Delhi, for their assistance in developing this plasma device.     
\section{Author Declarations}
\subsection{Conflict of Interest}
The authors declare that they have no conflicts of interest in connection with this article.
\subsection{Author Contributions}
Dr Choudhary conceived the idea for this project and developed the experimental setup. Vanshika performed the experiments on dye degradation. Mr Surya assisted in experiments and performed spectroscopy of dye solutions. The first draft of the manuscript was written by Mangilal Choudhary and revised by all authors. All the authors read and approved the final manuscript.   
\section{Data Availability}
The datasets generated during and/or analyzed during the current study are available from the corresponding author upon reasonable request.
\bibliographystyle{jpp}
\bibliography{biblography}
\end{document}